\documentclass[usenatbib]{mn2e}

\makeatother
\usepackage[latin1]{inputenc}
\usepackage{multirow}
\usepackage{graphicx}
\usepackage{multicol}
\usepackage{float}
\usepackage{color}
\usepackage{multirow}
\usepackage{footnote}
\usepackage{fancyheadings}
\usepackage{amssymb}
\usepackage[figuresright]{rotating}
\usepackage{amsmath}
\usepackage{url}
\title[Deep H$\alpha$ Observations of NGC 253]{Deep H$\alpha$ Observations of NGC 253: a Very Extended and Possibly Declining Rotation Curve?}
\author[J. Hlavacek-Larrondo et al.]{J. Hlavacek-Larrondo$^{1,2}$\thanks{E-mail: juliehl@ast.cam.ac.uk}, C. Carignan$^{1,3}$, O. Daigle$^{1,4}$, M.-M. de Denus--Baillargeon$^{1}$, 
\newauthor{M. Marcelin$^{4}$, B. Epinat$^{4,5}$, O. Hernandez$^{1}$} \\ 
{\rm $^1$Laboratoire d'Astrophysique Expérimentale, Département de physique, Université de Montréal, C.P. 6128, Succ. centre-ville, Montréal,} \\ {\rm Québec, Canada, H3C 3J7}\\ 
{\rm $^2$Institute of Astronomy, University of Cambridge, Madingley Road, Cambridge CB3 0HA}\\
{\rm $^{3}$Observatoire d'Astrophysique de l'Université de Ouagadougou, BP 7021, Ouagadougou 03, Burkina Faso} \\ 
{\rm $^{4}$Laboratoire d'Astrophysique de Marseille, Université de Provence, CNRS, 38 rue Frédéric Joliot-Curie, F-13388 Marseille Cedex 13,} \\ {\rm France} \\ 
{\rm $^{5}$Laboratoire d'Astrophysique de Toulouse-Tarbes, Université de Toulouse, CNRS, 14 Avenue Edouard Belin, F-31400 Toulouse, France} }

\begin{document}

\date{Accepted 2010 September 6. }

\pagerange{\pageref{firstpage}--\pageref{lastpage}} \pubyear{2009}

\maketitle

\label{firstpage}

\begin{abstract}
This study presents a deep H$\alpha$ kinematical analysis of the Sculptor Group galaxy NGC253. The Fabry-Perot data were taken with the 36-cm Marseille Telescope in La Silla, Chile, using an EMCCD detector. Typical emission measures of $\sim$ 0.1 cm$^{-6}$ pc are reached. The observations allow the detection of the Diffuse Ionized Gas component through [N\thinspace{\sc ii}] emission at very large radii of 11.5$'$, 12.8$'$ and 19.0$'$, on the receding side of the galaxy. No H$\alpha$ emission is observed at radii larger than the neutral component (11.5$'$). The very extended rotation curve confirms previous results and shows signs of a significant decline, on the order of 30 per cent $v_\mathrm{max}$. Using the rotation data, mass models are constructed with and without the outer [N\thinspace{\sc ii}] data points, and similar results are found. The declining part of the rotation curve is very well modeled, and seems to be truly declining. 
\end{abstract}


\section{\large INTRODUCTION}\label{a1c1}

The Sculptor Group harbours an impressive amount of gas-rich late-type galaxies \citep{deV1959130}, including NGC 7793, NGC253, NGC 300, NGC 247 and NGC 55. Several other small galaxies are known to be gravitationally bound to the group \citep{Cot1997114}. The galaxies are located at a distance estimated between 1.5 and 4 Mpc (Jerjen, Freeman \& Binggeli 1998\nocite{Jer1998116}; Karachentsev et al. 2003\nocite{Kar2003404}). Its proximity, as well as the fact that its galaxies are isolated, makes it an ideal ground for studying the kinematical properties of disc galaxies. Moreover, the majority of its galaxies being late-type with no or almost no bulge, renders them excellent tools for studying the properties of dark matter haloes \citep{Car1985294}.

The literature contains many studies of the Sculptor Group galaxies. Among them, a deep H$\alpha$ study of NGC 7793 was performed by \citet{Dic2008135}. This study, based on deep optical Fabry-Perot (FP) data, confirmed that NGC 7793 has a truly declining rotation curve. Deep surveys of galaxies are essential to extract very extended rotation curves, and allow better constraints on the dark halo parameters. This study aims at studying the dynamical properties of another galaxy belonging to the Sculptor Group, NGC 253, also using deep H$\alpha$ FP data. 

NGC 253 is a late-type spiral starburst galaxy. The central regions show significant activity, and have therefore been thoroughly studied. H53$\alpha$ and H72$\alpha$ observations showed the presence of a counter-rotating central disc \citep{Zha2001240}. Furthermore, observations of the H$_2$ emission line suggest the presence of gas being ejected to distances of 130 parsecs above the disc \citep*{Sug2003584}. 

\citet*{Puc1991101} presented a kinematical study of the neutral gas component (H\thinspace{\sc i}) using 21-cm synthesis observations. A rotation curve extending to a maximum radius of 8.5 kpc (or 11.5$'$) was derived, and showed that even at the largest radius, the rotation curve continues to rise with a maximum velocity of 224 km s$^{-1}$. A long-slit H$\alpha$ study has also been done by \citet{Arn1995110}. 

The Diffuse Ionized Gas (DIG) is an important component of the interstellar medium that can extend to large radii. Studies have shown that the gas is both warm, $T\sim6000-10000K$ \citep{Rey1992392}, and diffused, $n_\mathrm{e}\sim0.03$ cm$^{-3}$ \citep{Dop2003}. However, the origin of the DIG still remains controversial. One of the leading theories implies ionizing photons escaping H\thinspace\thinspace{\sc ii} regions and reaching very large distances of several kiloparsecs. \citet*{Hoo1996112} studied the properties of this diffuse component in NGC 253. They measured the fraction of H$\alpha$ luminosity from the DIG, and found it to be between 35 and 43 per cent. The distribution of the DIG also seemed to be concentrated near the H\thinspace{\sc ii} regions, extending away from them up to a maximum radius of 1 kpc. 

\begin{table}
\centering
\caption{Optical parameters of NGC 253 \label{a1_t1}}
\resizebox{0.46\textwidth}{!}{%
\begin{tabular}{@{}llllllllllllll@{}}
\hline
\hline
Morphological type$^{2}$ & SAB(s)c \\
R.A (2000)$^{1}$ & 00$^h$47$^m$33.1$^s$ \\
Dec. (2000)$^{1}$ & -25$^o$17$^{'}$18${''}$ \\
Isophotal major diameter$^{1}$, $D_\mathrm{25}$ (\textit{B}) & 27.5${'}$ \\
Holmberg radius$^{2}$, $R_\mathrm{HO}$ (\textit{B}) & 17.5${'}$ \\
Exponential scale length$^{2}$, $\alpha^{-1}$ (\textit{B}) & 2.4 kpc \\
Axis ratio$^{2}$ ($q\equiv b/a$) & 0.23 \\
Inclination$^{2}$ ($q_\mathrm{0}=0.15$), \textit{i} & 79.8$^{o}$ \\
Total apparent \textit{B} magnitude, $B^{0,i}_\mathrm{T}$$^{5}$& 8.05 \\
Total corrected apparent \textit{B} magnitude, $B^{0,i}_\mathrm{T}$$^{3,4}$& 7.09 \\
Adopted distance (Mpc)$^{6}$ & 2.58 \\
 & ($1^{'}=0.75$ kpc)\\
Absolute \textit{B} magnitude, M$^{0,i}_\mathrm{B}$ & -19.97 \\
Total blue luminosity (M$_\odot=5.43$), $L_\mathrm{B}$$_\odot$ & $14.4\times10^{9}$ \\
\hline
$^1$RC3 & \\
$^2$\citet{Puc1991101} & \\
$^3$Internal absorption A$(i)=0.91$ (RC3). & \\
$^4$Galactic extinction A$_\mathrm{B}=0.05$ (RC3). & \\
$^5$\citet{Pen1980239} & \\
$^6$\citet{Puc198895} & \\
\end{tabular}}
\label{a1_t1}
\end{table}
The kinematical properties of the DIG in NGC 253 were studied by \citet*{Bla1997490}. They found very extended H$\alpha$ and [N\thinspace{\sc ii}] emission on the south-west side of the galaxy at radii larger than that of the neutral component (H\thinspace{\sc i}). The neutral component is detected out to 11.5$'$, while the optical component extends to 15.0$'$. The last observational point of the optical rotation curve also suggests a decreasing rotation curve. Observing emission at these radii could therefore have great implications, since it would allow the study of the distribution of the dark halo component at very large radii, and perhaps even allow the study of the entire distribution of matter. 

This paper presents deep optical FP data of the Sculptor Group galaxy NGC 253. A look for faint emission on both sides of the galaxy will be performed, as well as a search for emission beyond a radius of 15.0$'$ in order to confirm the declining rotation curve seen by \citet{Bla1997490}. Section 2 will give a description of the FP observations and of the data reduction techniques used to extract the velocity field and the total integrated H$\alpha$ emission map. It will also describe a method developed for the flux calibration of the H$\alpha$ FP data. Section 3 shows the velocity field obtained in this study, and Section 4 presents the extracted kinematical parameters and adopted rotation curve. The mass models analysis is described in Section 5. The results are then analysed and discussed in Section 6 and Section 7 presents a summary and the conclusions. The optical parameters of NGC 253 are summarized in Table \ref{a1_t1}. A distance of 2.58 Mpc has been adopted for this study \citep{Puc198895}.

\section{\large{Fabry-Perot observations and data reduction}}\label{a1c2}

\subsection{\normalsize Observations}\label{a1c21}

The FP observations were obtained on the 36-cm Marseille Telescope at La Silla, Chile, from 2007 October 3 to 2007 October 15. A commercial camera Andor iXon, equipped with an EMCCD detector, was used \citep[see][]{Dai20045499}. All the observations were made with the same filter, camera and FP etalon. Dark frames were taken at the beginning and end of each night. Furthermore, data were taken to calibrate the gain of the EMCCD. Flats with the etalon and H$\alpha$ filter kept in place were also taken at twilight. The interference filter had a central wavelength of $\lambda_\mathrm{c}=6561.0$ \AA , a FWHM of 30 \AA , and a maximum transmission of $\sim80$ per cent at a temperature of 10$^\circ$ Celsius, see Fig. \ref{a1f3add}. The FP interference order was $p=765$ at H$\alpha$, with a Free Spectral Range (FSR) of 8.58 \AA . The mean \textit{Finesse} during the observations was $\sim$16, with an etalon spectral resolution of $\sim0.53$ \AA\ ($\Delta\lambda=\frac{FSR(\AA )}{Finesse}$). The data were scanned throughout the FSR over 40 channels of 0.21 \AA\ per channel. Wavelength calibration was performed using a neon lamp (Ne) at 6598.95 \AA . Several observations of the calibration cube were made during each night. For each observation of the calibration data, the etalon was scanned during one complete cycle (40 channels), with an exposure of 1 second per channel. The EMCCD has $512\times512$ pixels and a resolution of 2.77$''$/pixel. Each field covers $\sim23.6'\times23.6'$. 

The EMCCD has a quantum efficiency of $\sim90$ per cent at H$\alpha$ and the Electron Multiplying gain used yielded a read-out noise of $\sigma\simeq0.05$ electron. In the Photon Counting Mode, the camera is operated at a higher frame rate and only one photon per pixel can be detected, which eliminates the excess noise factor. Nonetheless, in order to lower the effect of the Clock Induced Charges generated by the Andor camera, the observations were done in Amplified Mode (15 s per image), where more than one photon per pixel per frame can be counted. This mode of operation causes the excess noise factor of the EMCCD to affect the photometric quality of the data. More details can be found in \citet*{Dai20066276} and \citet{Dai20045499,Dai20087014}. 

NGC 253 was observed over three fields: centre, north-east and south-west. See Fig. \ref{a1f4} for outlines of the fields. Each field was exposed with an exposure time of 15 seconds per channel. The central field was exposed for 64 cycles, corresponding to 640 minutes or 10.7 hours. The north-east and south-west fields were each exposed for 38 cycles, corresponding to 380 minutes or 6.3 hours. 

\subsection{\normalsize Data reduction}\label{a1c23}

\subsubsection{\normalsize Reduction steps}\label{a1c231}

Raw observational data consist of many data files taken for every cycle and channel. Each observational run consists of one to twelve cycles, and for each run, a wavelength-sorted data cube can be extracted. The first step is to correct the data with the normalized flat image. In order to correct for the airmass dependance, which can vary considerably during an observing run, the correction must be applied prior to the creation of a wavelength-sorted data cube. These two corrections are applied at the beginning of the reduction procedure, for each and every channel of every cycle. The following steps were adopted for the reduction of the data: 

1. Correction of all data with the normalized flat image. 

2. Creation of a wavelength-sorted data cube corrected for airmass dependance, guiding shifts, cosmic rays, dark, and gain.

3. Ghosts correction\footnote[1]{ Ghosts are reflections of bright H\thinspace{\sc ii} regions in the field. Spatially, the ghosts are located symmetrically opposite with respect to the optical axis and they will be at the same wavelength as the primary image. They can thus be well identified, both in position and in velocity. A ghost correction was applied using the method developed by \citet{Epi2008388}. Some residuals of ghosts were seen, but these fell far away from the galaxy. They also did not fall in the pie-shape regions used to study the DIG (see Section 3 and 4).} \citep[see][]{Epi2008388}.
 
4. Heliocentric velocity correction. 

5. Hanning spectral smoothing and sky subtraction \citep[see][]{Dai20066276}. 

The data cubes were then summed for each field of NGC 253, using the position of bright stars. The three fields were then mosaicked together, and corrected for the different exposure times using the computed flux of bright stars in each field. The final data cube has $1024\times512$ pixels. 

6. Spatial $5\times5$ binning procedure. 

7. Single gaussian fitting procedure. 

8. Correction for the filter dependance: division of the profile by the filter spectrum. 

9. Extraction of the velocity, velocity dispersion and H$\alpha$ integrated maps. 

\subsubsection{\normalsize Lorentzian and gaussian profiles}\label{a1c232}

The intensity weighted method, which is often used to extract the velocity of galaxies through spectral-line profiles, calculates the barycentre of the spectral line, and is appropriate for cases of low S/N data \citep[see][for more details]{Amr1991}. However, the shape of the emission lines observed should be the result of a convolution between the device function, an Airy function, and the emitted spectral line, which is assumed to be gaussian. Thermal agitation of the gas dominates, and is responsible for the gaussian shape of the profile. The observed profile has therefore been approximated by a Lorentzian or a gaussian profile. The gaussian profile yielded a mean reduced ${\chi^2}_\mathrm{r}$ of 1.37 for the entire galaxy, while the Lorentz profile yielded a mean ${\chi^2}_\mathrm{r}$ of 2.67. The gaussian fitting procedure was therefore adopted. 

\subsection{\normalsize H$\alpha$ flux calibration}\label{a1c22}

The H$\alpha$ flux calibration of the data was done during the reduction process. In order to carry out the calibration, an extinction coefficient as well as a conversion factor (\textit{FAC}) must be determined as defined by Eq. \ref{a1_eq1} and Eq. \ref{a1_eq2}, where \textit{Flux} is the corresponding flux value, \textit{X} is the airmass at the time of exposure, \textit{C$_\mathrm{obs}$} is the observed number of counts, and \textit{C$_\mathrm{true}$} is the true number of counts. In order to alleviate the total number of parameters to adjust, the zero point parameter, \textit{B}, and the initial conversion factor, \textit{f$_\mathrm{ac}$}, were combined together to obtain the conversion factor (\textit{FAC}). 
\begin{eqnarray}
Flux & = & f_\mathrm{ac} \cdot C_\mathrm{true} \nonumber \\ & = & f_\mathrm{ac} \cdot 10^{(A \cdot X + B)/2.5} \cdot C_\mathrm{obs} \nonumber \\ & = & 10^{(A \cdot X)/2.5} \cdot C_\mathrm{obs} \cdot f_\mathrm{ac} \cdot 10^{(B/2.5)} 
\label{a1_eq1}
\end{eqnarray}
\begin{eqnarray}
Flux & = & 10^{(A \cdot X)/2.5} \cdot C_\mathrm{obs} \cdot (FAC)
\label{a1_eq2}
\end{eqnarray}
For this study, the value of the extinction coefficient published by ESO was used. The value for H$\alpha$ at the La Silla site is 0.07 mag/airmass \citep[see also][]{Bur1995112}. 

\begin{figure}
\centering
\begin{minipage}[c]{0.90\linewidth}
 \centering \includegraphics[width=\linewidth]{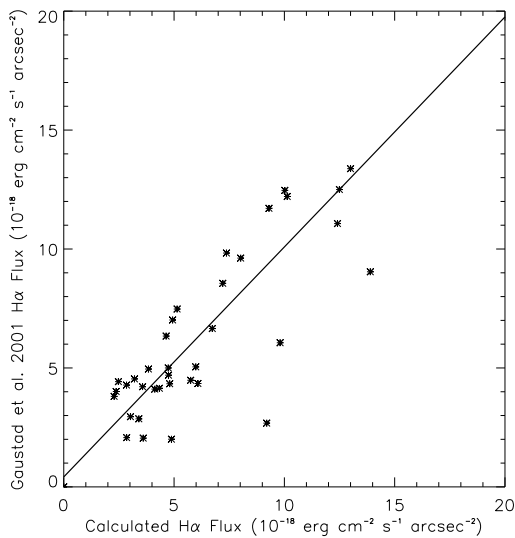}
\end{minipage}
\caption{Comparison between the fluxes obtained for NGC 253 in this study, and those of \citet{Gau2001113}. Each point represents a bright H\thinspace{\sc ii} region in NGC 253. The conversion factor was determined to be 4576 (1 count = 4576 dR, where 1/6R or an EM of 2/6 cm$^{-6}$ pc is equivalent to $10^{-18}$ ergs cm$^{-2}$ s$^{-1}$ arcsec$^{-2}$). The continuous line has a slope set to unity.}
\label{a1f1}
\end{figure}

The next step was to extract the conversion factor between the number of counts corrected for airmass and the corresponding flux. This was accomplished once all the reduction steps were performed, and the total number of counts was obtained, since it allows a better constraint on the factor determined. The continuum subtracted H$\alpha$ map from the SHASSA (Southern H$\alpha$ Sky Survey) catalogue was used to flux calibrate the data \citep[][http://amundsen.swarthmore.edu/]{Gau2001113}. This catalogue has units of deci-Rayleigh, dR (1R = $10^6$ photons cm$^{-2}$ s$^{-1}$ sr$^{-1}$ or $0.6\times10^{-17}$ ergs cm$^{-2}$ s$^{-1}$ arcsec$^{-2}$ or to an Emission Measure, EM, of $\sim2$ cm$^{-6}$ pc). Moreover, the SHASSA image was corrected for [N\thinspace{\sc ii}] ($\lambda\lambda$ 6548 \AA , 6583 \AA ) contamination. For NGC 253, the intensity of the contamination was considered to be 0.48 times that of the H$\alpha$ intensity \citep{Ken2008178}. Afterwards, a linear regression was applied between the counts observed from this study and those of the SHASSA catalogue for individual bright H\thinspace{\sc ii} regions in NGC 253. Fig. \ref{a1f1} compares the calibrated data from this study with those of the SHASSA image for bright H\thinspace{\sc ii} regions. 

\section{\large{Velocity field}}\label{a1c3}

\begin{figure*}
\centering
\begin{minipage}[c]{0.47\linewidth}
 \centering \includegraphics[width=\linewidth]{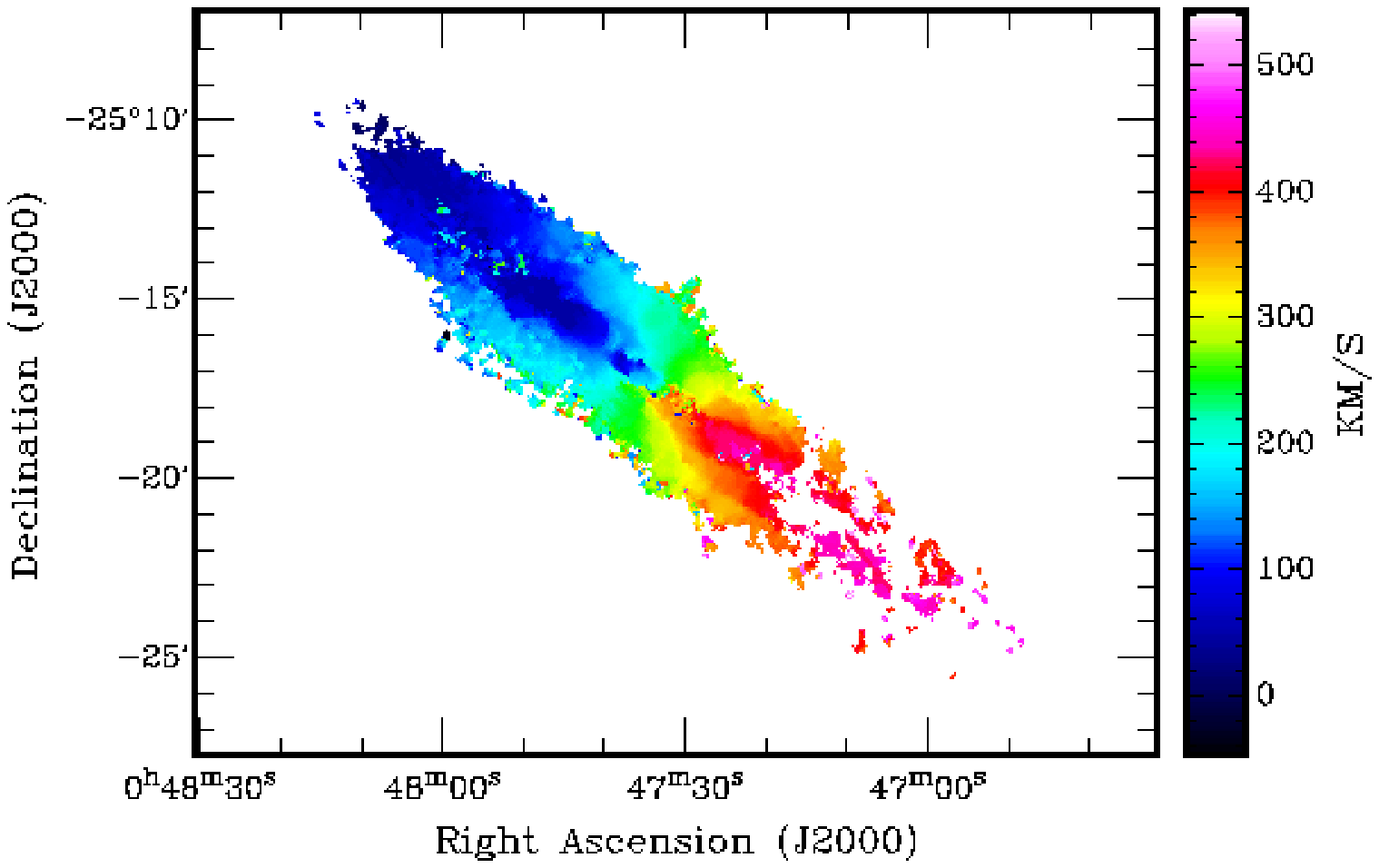}
\end{minipage}
\begin{minipage}[c]{0.47\linewidth}
\vspace{-5pt}
 \centering \includegraphics[width=\linewidth]{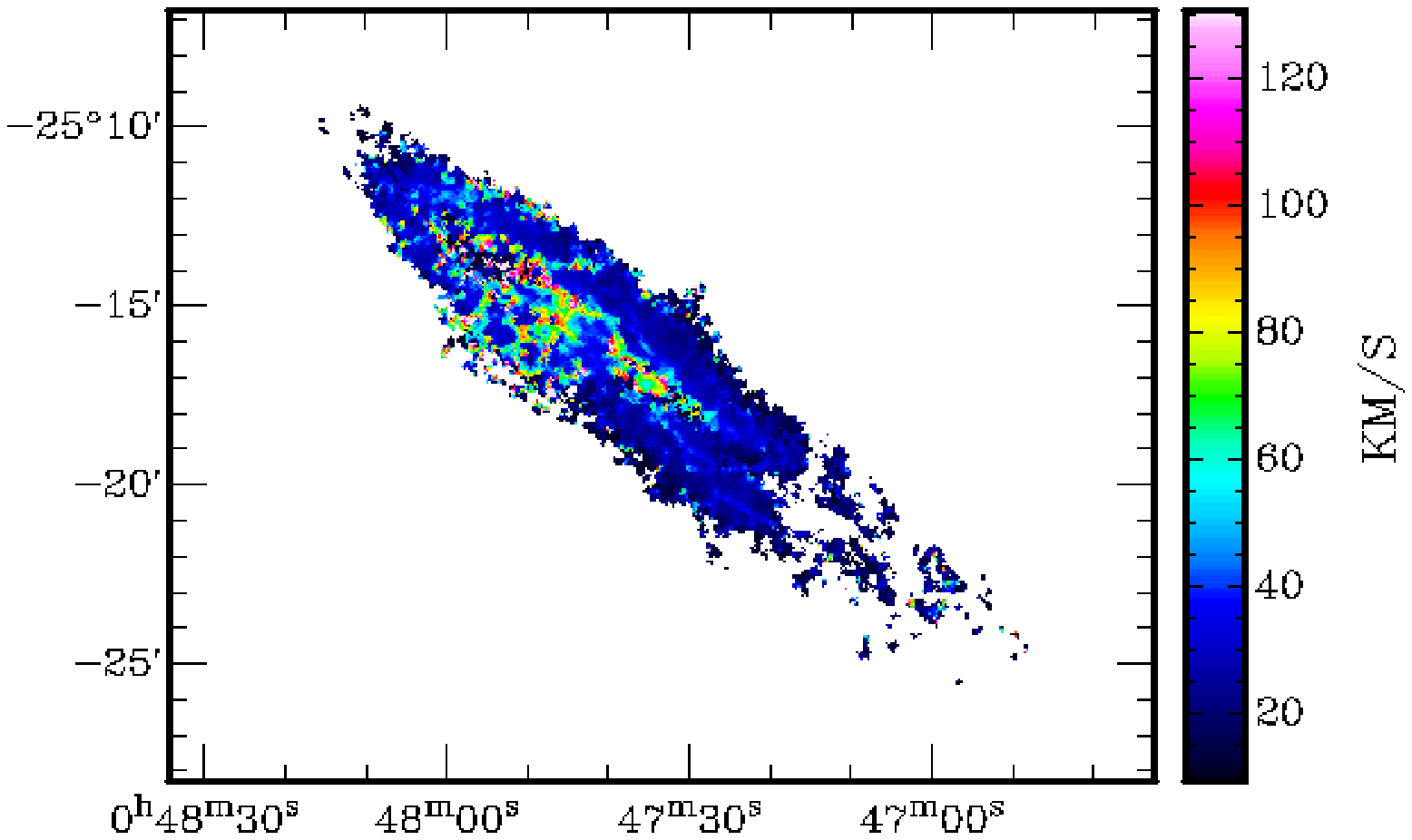}
\end{minipage}
\begin{minipage}[c]{0.48\linewidth}
 \centering \includegraphics[width=\linewidth]{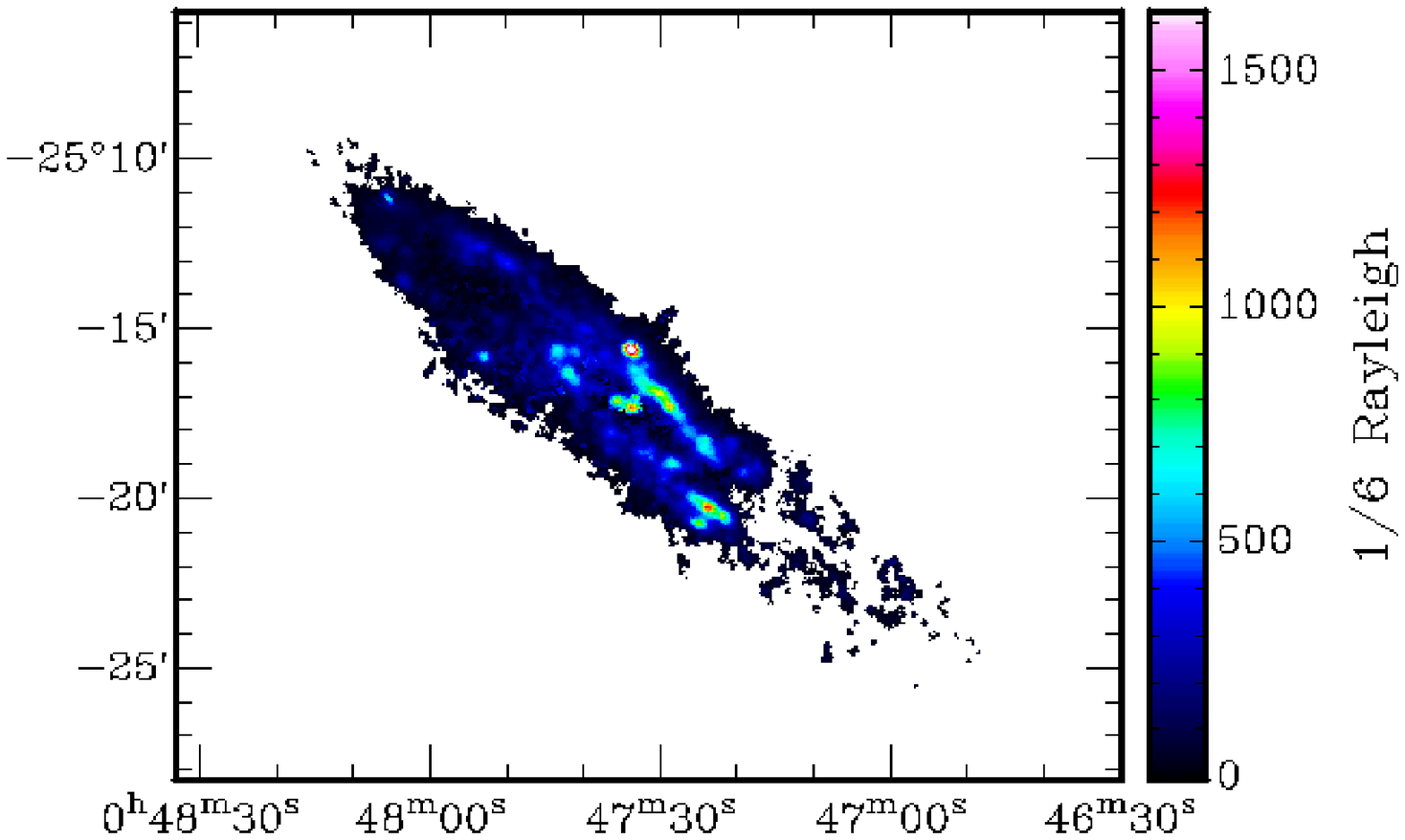}
\end{minipage}
\begin{minipage}[c]{0.47\linewidth}
 \centering \includegraphics[width=\linewidth]{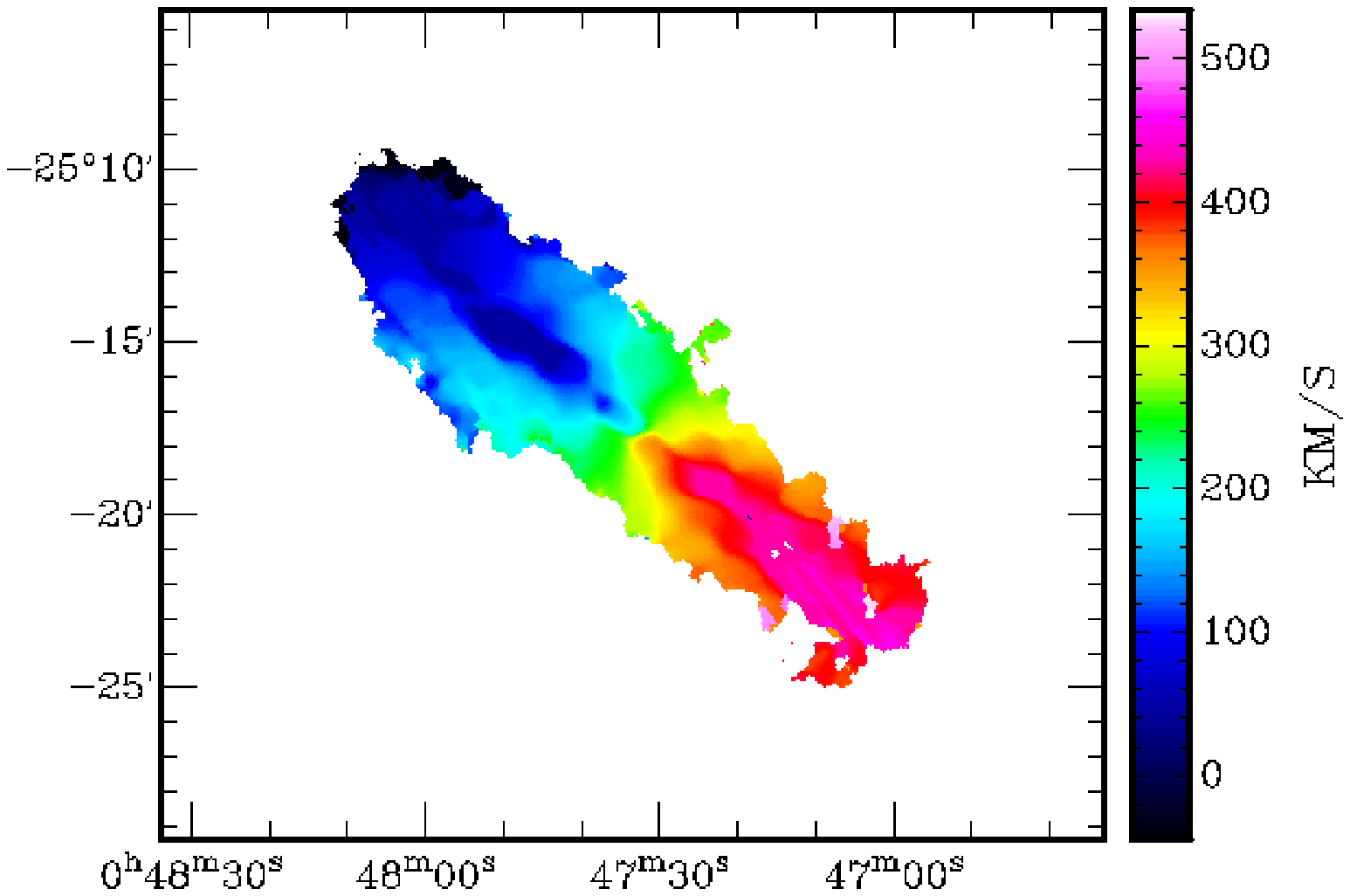}
\end{minipage}
\caption{Kinematical maps of NGC 253 obtained after a 5 pixels $\times$ 5 pixels square binning and a single Gaussian fitting procedure, unless otherwise specified. Top-left: H$\alpha$ velocity field. Top-right: Dispersion map. Bottom-left: Flux calibrated integrated map obtained after correction for [N\thinspace{\sc ii}] contamination (scaled in 1/6 Rayleigh or $10^{-18}$ ergs cm$^{-2}$ s$^{-1}$ arcsec$^{-2}$). Bottom-right: H$\alpha$ velocity field with a gaussian spatial smoothing of $w_\lambda=55''$.}
\label{a1f3}
\end{figure*}

\begin{figure*}
\centering
\begin{minipage}[c]{0.5\linewidth}
 \centering \includegraphics[width=\linewidth]{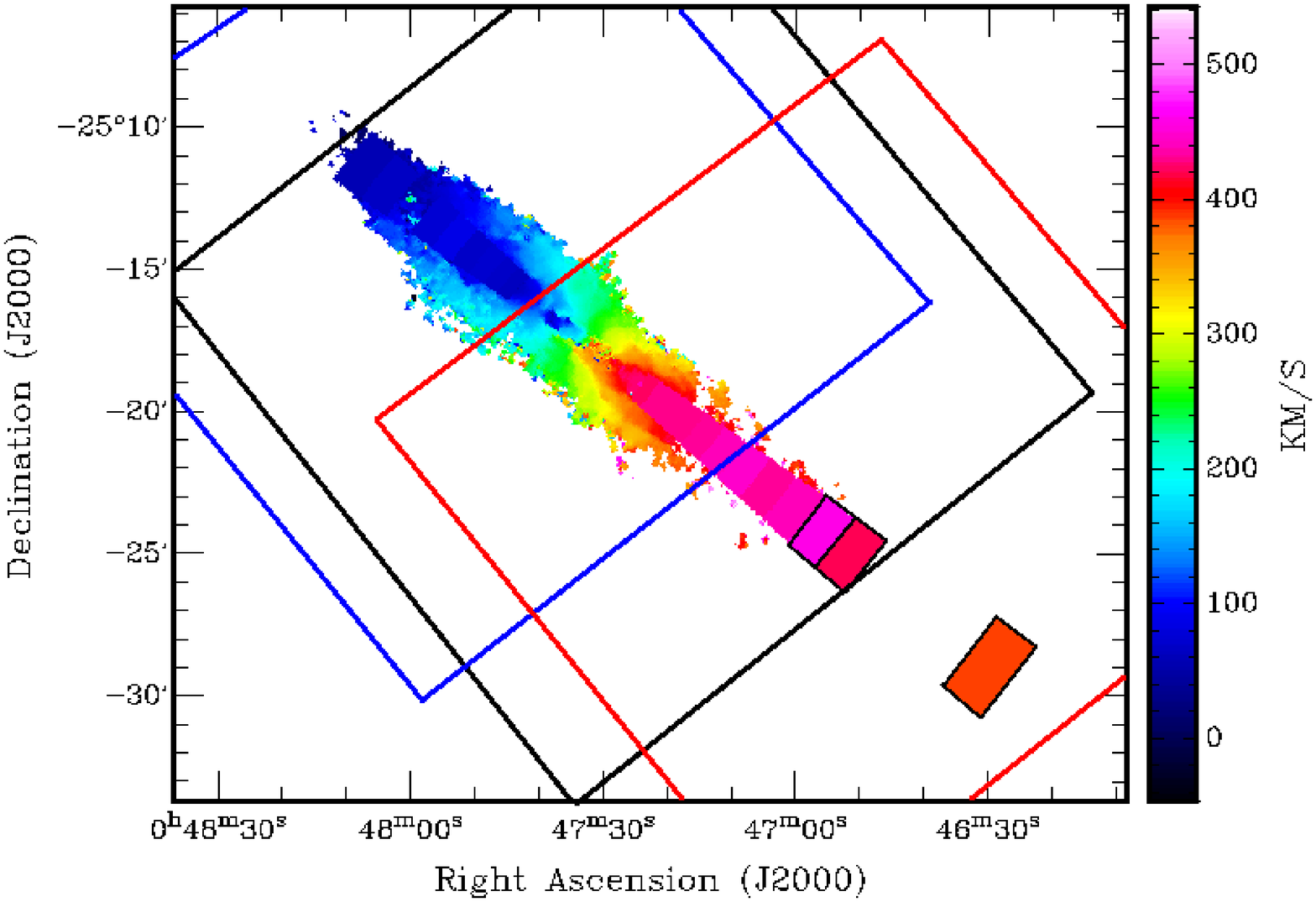}
\end{minipage}
\begin{minipage}[c]{0.4\linewidth}
\vspace{-23pt}
 \centering \includegraphics[angle=270.,width=\linewidth]{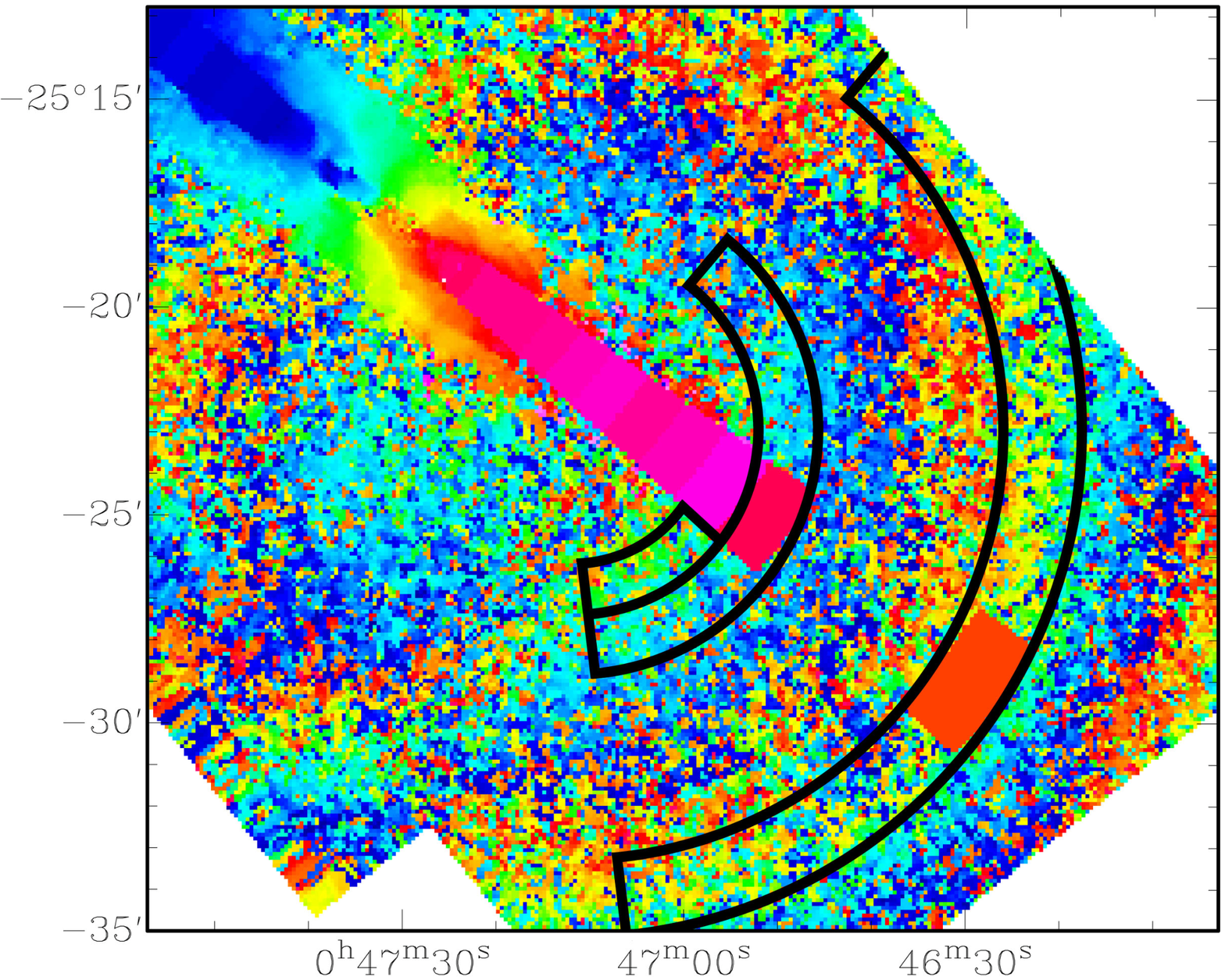}
\end{minipage}
\begin{minipage}[c]{0.31\linewidth}
 \centering \includegraphics[width=\linewidth]{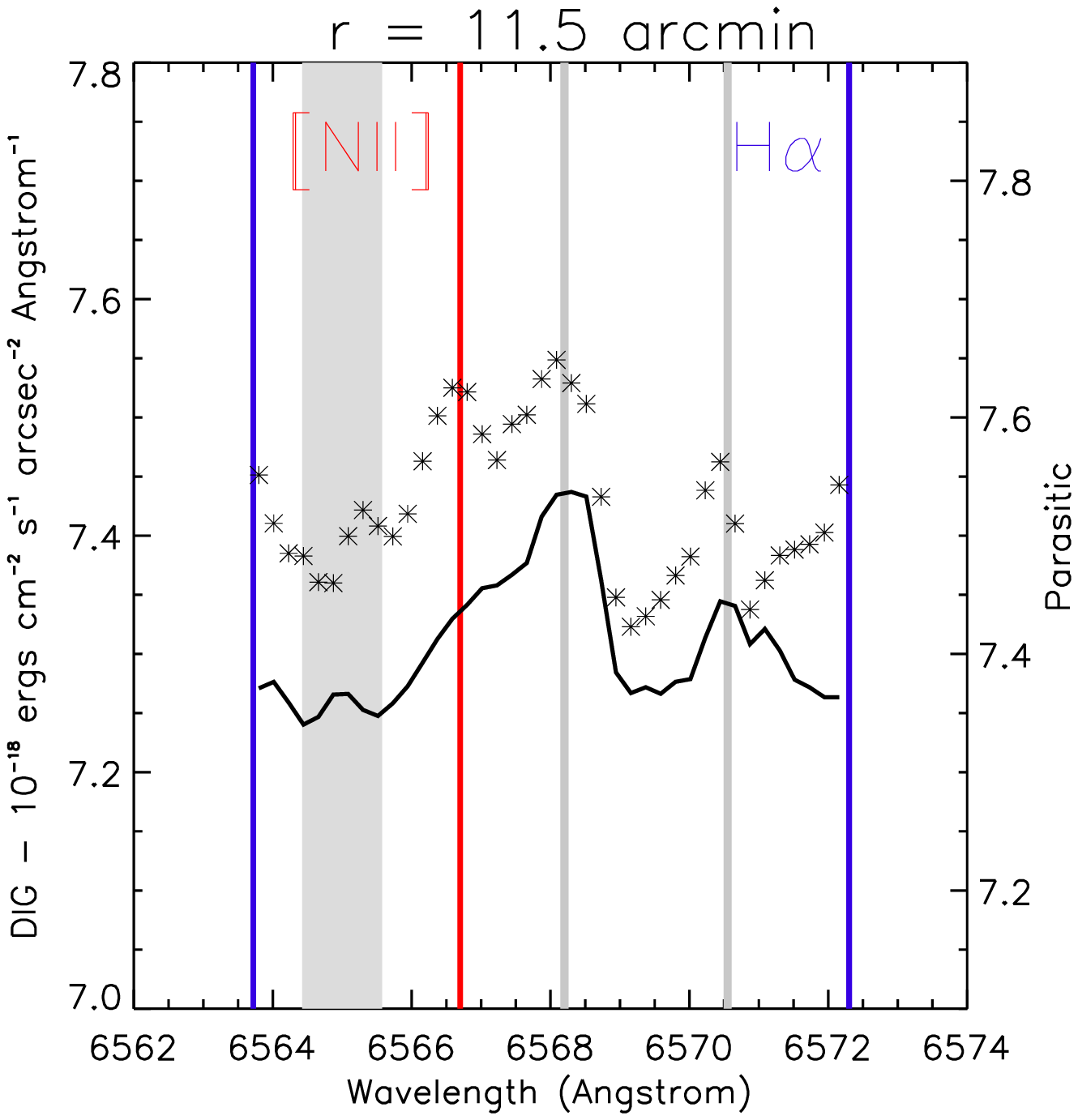}
\end{minipage}
\begin{minipage}[c]{0.31\linewidth}
 \centering \includegraphics[width=\linewidth]{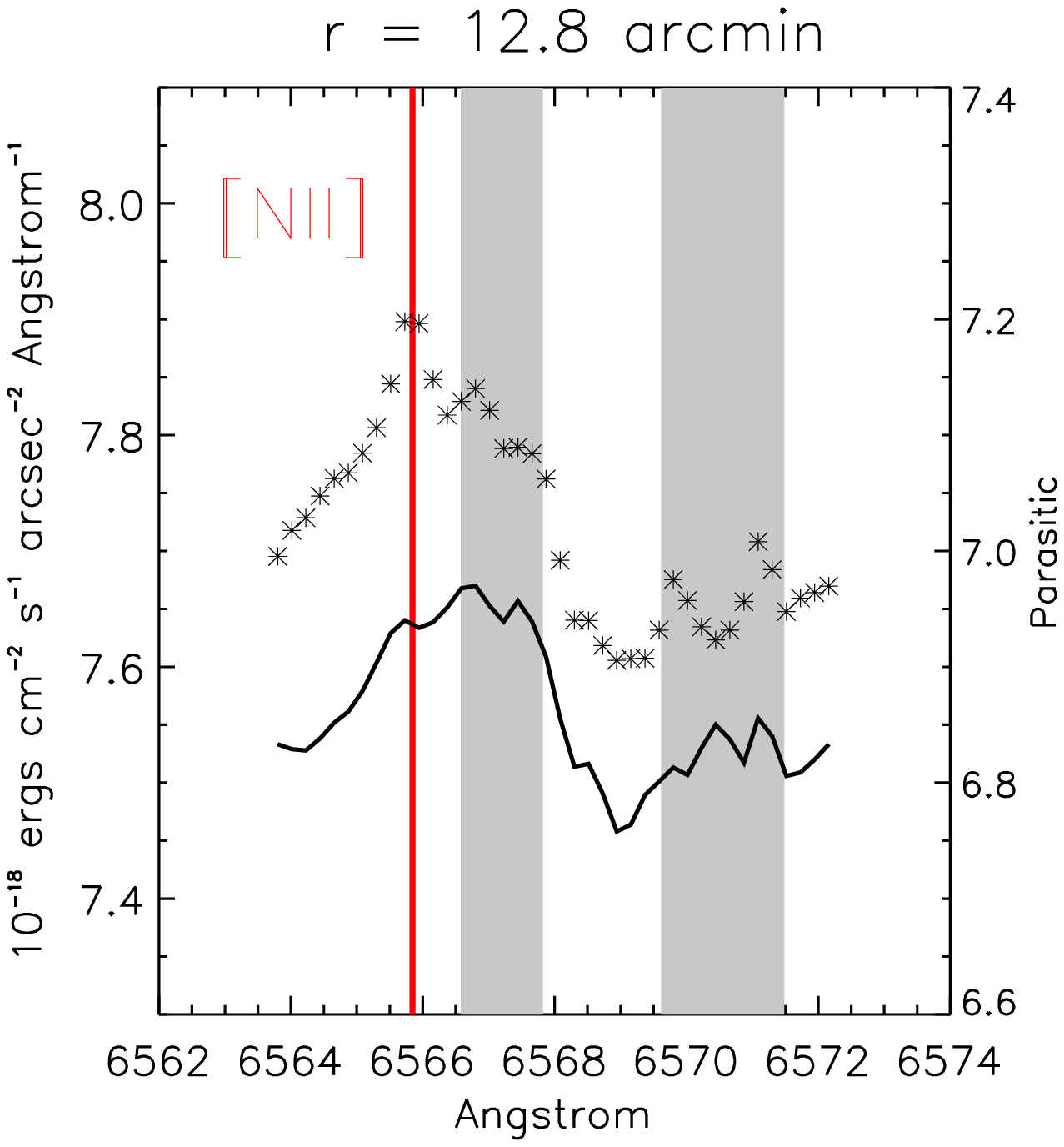}
\end{minipage}
\begin{minipage}[c]{0.31\linewidth}
 \centering \includegraphics[width=\linewidth]{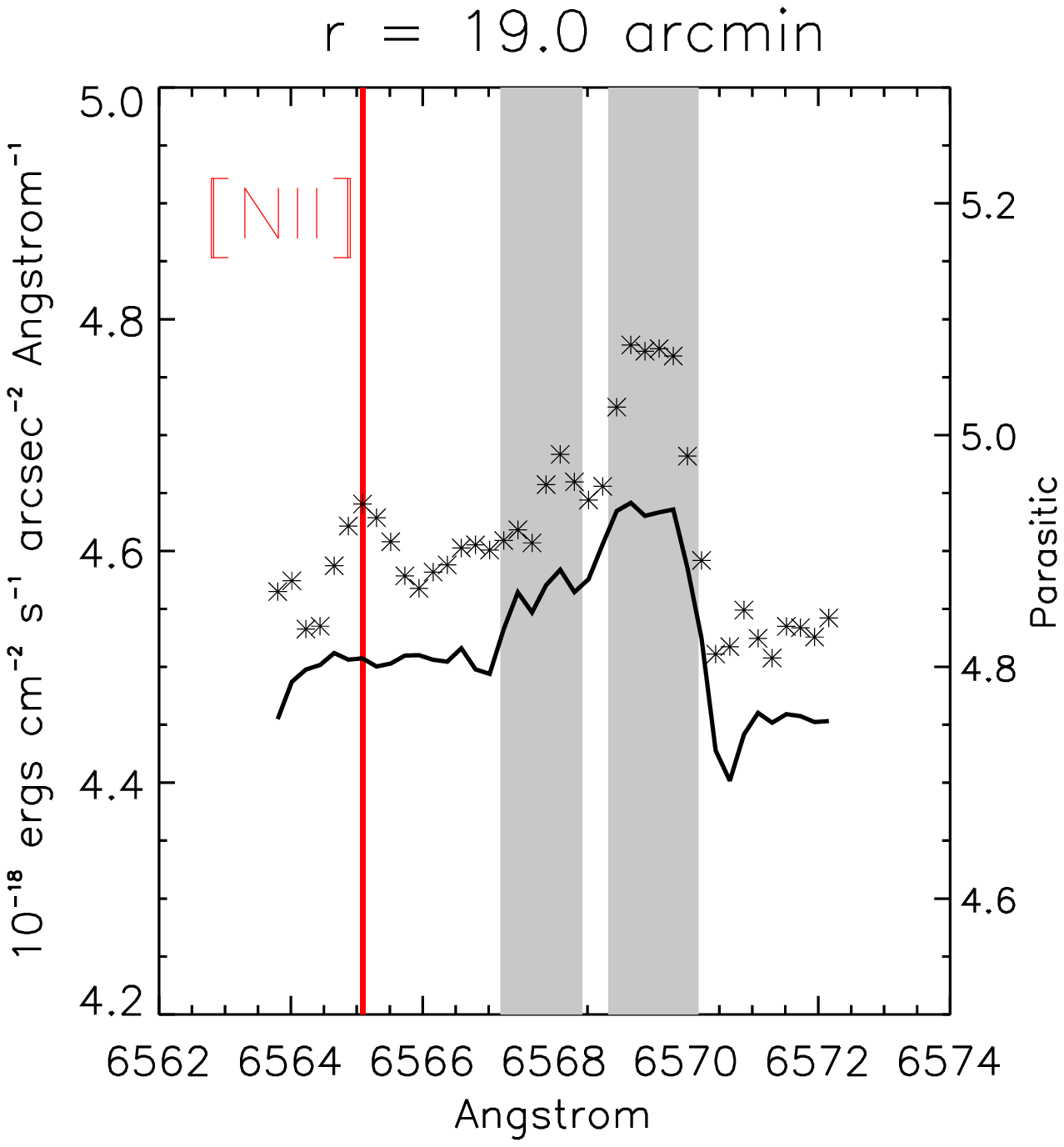}
\end{minipage}
\caption{Top-left: Velocity field of NGC 253 obtained after a $5\times5$ square binning procedure, a pie-shaped binning procedure, and a single gaussian adjustment to the emission profiles. The last three binned intervals (black contours) are the result of [N\thinspace{\sc ii}] emission. Outlines of the three fields are also plotted in large squares: black for centre field, blue for north-east field and red for south-west field. Top-right: Close up of same velocity field, but without any masking applied - the rings of the parasitic emission embedded in the velocity field are clearly seen. The black annuli contours show the regions binned to extract the typical parasitic spectra. Bottom: Spectra of the last three binned intervals of the pie-shape in order of radius (from left to right). The star-like points are for the spectra of the DIG and the continuous black profile are for the corresponding typical parasitic spectra. The blue vertical line represents the location of the H$\alpha$ line, the red vertical lines represent those of the [N\thinspace{\sc ii}] emission, and the grey regions give an indication of the location of the contaminated areas.}
\label{a1f4}
\end{figure*}

The velocity field was extracted using three distinct methods. The first followed the reduction steps outlined in Section 2.3.1, and the resulting velocity field is presented in the top-left image of Fig. \ref{a1f3}. The corresponding velocity dispersion map and H$\alpha$ integrated map are also shown. As expected from an inspection of the velocity dispersion map, the velocity field contains perturbations, notably in the inter-arm region of the northern side. In order to determine the global kinematical parameters, a second method using a gaussian smoothing with $w_\mathrm{\lambda}=55''$ (20 pixels) was used to extract the velocity field, where \textit{w}$_\mathrm{\lambda}$ is the width in $''$ of the smoothing function (see bottom-right image of Fig. \ref{a1f3}). 
In order to study the very diffuse emission at extended radii, a third method was used to extract the velocity field. A special binning procedure similar to that used by \citet{Bla1997490} was applied. The smoothing function has the shape of an isoscele triangle (pie-shape), centred around the major axis of the galaxy and about its dynamical center. For a given opening angle, different regions corresponding to different radii intervals were binned together. An opening angle of $\sim15^o$ centred around the major axis was chosen, with the binning intervals growing with radius (see Fig. \ref{a1f4}). 

The central regions of NGC 253 are very bright. It is therefore possible that this bright source causes the appearance of parasitic emission in the field, similar to ghosts. The light from the nucleus is a strong continuum source. When passing through the filter, diffusion of the light can be observed. These ghosts will appear as rings in the velocity map, on top of the velocity field of the galaxy. The following section describes in detail this parasitic emission, as well as the method used to distinguish it from the diffuse emission of the galaxy.

\section{\large{Isolating the extended diffuse emission}}

\subsection{Diffuse Ionised Gas}

The term DIG refers to a component of the interstellar medium that is both warm, $T\sim6000-10000K$ \citep{Rey1992392}, and diffused, $n_\mathrm{e}\sim0.03$ cm$^{-3}$ \citep{Dop2003}. The DIG has low values of EM, since the emission measure depends on the square of the density. This component can be studied by separating its emission from the bright, high EM, ionised regions in a galaxy, such as H\thinspace{\sc ii} regions. Some have simply separated this component by applying a cut off: any emission below a certain EM, such as $\sim80$ pc cm$^{-6}$ \citep[value taken from][]{Fer1996111}, was considered to be from the DIG. 

However, \citet{Hoo1996112} argue that this simple cut off method is not adequate for highly inclined galaxies such as NGC 253, and that it can result in underestimating the total DIG component. In their study, they developed a special median filtered masking technique to isolate the DIG emission from their NGC 253 images, see \citet{Hoo1996112} for more details. 

Nonetheless, the goal of this study is not to separate the DIG emission from the bright ionised regions. The goal is to obtain a very deep image of the galaxy in order to study the kinematics at very extended radii. This does not require isolating the DIG component, since any emission whether with high or low EM can be used to obtain kinematical measures. 

For the extended ionised emission detected beyond $\sim11'$, we use the term DIG. It is expected that at large radii, the EM of the gas drops significantly. For radii beyond $\sim11'$ in NGC 253, the EMs are well below a few pc cm$^{-6}$, which qualifies it has DIG regardless of the inclination of the galaxy. In the following section, we explain how this very low EM gas was detected, and how its emission was isolated from the parasitic emission mentioned in Section 3. 

\subsection{Isolating the extended DIG emission from the parasitic emission}

The parasitic emission seen in the data of this paper is thought to originate from diffusion of the bright nucleus light when passing through the filter. The same parasitic emission was seen on other observing runs, when the targets were bright point-like planetary nebulae. Since the nucleus of NGC 253 is particularly bright and point-like, the parasitic emission seen could therefore have the same cause which is thought to be diffusion of light through the filter. 

The top-right panel of Fig. \ref{a1f4} shows a close up of the raw velocity field (no masking applied), where the parasitic emission is clearly seen as concentric circular rings embedded in the velocity field of NGC 253. The centre of the parasitic rings coincides with the FP optical centre, and so rings were seen for each field (centre, north-east, south-west). 

In Fig. \ref{a1f3}, the velocity field of the galaxy extends up to a radius of $\sim11'$. The intensity of the emission lines that gave forth to this velocity field are on average 6 times stronger than the intensity of the parasitic emission. Thus, for regions of strong H$\alpha$ emission, i.e. up to 11$'$, the parasitic emission is not problematic since the emission lines of the gas can be easily identified. The spectrum of each pixel in this velocity field was also individually examined, and all data that did not clearly show a strong emission line were discarded. 

For very low EMs, i.e. beyond a radius of 11$'$, the line intensity can be as low as the intensity of the parasitic emission. Diffuse emission can therefore be hidden in the signal of the parasitic emission. However, Fig. \ref{a1f4} shows that the parasitic emission is not random noise; it has a well defined spectra which gives circular rings in the velocity field. Therefore, a typical spectrum of the parasitic emission can be extracted for each radius interval, and used to identify if the emission seen is from the extended DIG or the parasitic emission. 

The method used to isolate the extended DIG emission is as follows, and is similar as the one used by Bland-Hawthorn et al. (1997) to identify the DIG emission in their study. This method is applied to all radii beyond 11$'$, where diffuse emission is seen. The DIG emission was extracted using a pie-shape binning procedure centred around the dynamical centre of the galaxy (see pie-shape in top-left of Fig. \ref{a1f4}). The spectra of the parasitic emission was extracted using another pie-shape binning procedure, this time centred around the centre of parasitic rings (see pie-shape outlined by black annuli in top-right of Fig. \ref{a1f4}). The opening angle was chosen to be large enough so that a typical signal for the parasitic emission could be extracted, while smoothing out any emission from the galaxy\footnote[2]{ This gave an opening angle of about $140^\circ$, except for the first radii bin ($r=11.5'\pm0.6'$) where a much smaller opening angle had to be chosen. An opening angle of $140^\circ$ would have overlapped significantly on emission from the galaxy.}. The radii intervals were chosen to be the same as the first binning procedure. Any emission profiles detected in the first procedure (pie-shape centred on galaxy) and not in the second (pie-shape centred on parasitic rings) should be the result of emission from the DIG. 

The bins with more than a $3\sigma$ detection are illustrated in the bottom panels of Fig. \ref{a1f4}. These all lie on the receding side of the galaxy. They are the last three intervals shown with black contours in the pie-shaped velocity field and are located at $11.5'\pm0.6'$, $12.8'\pm0.7'$, $19.0'\pm0.9'$ (top-left of Fig. \ref{a1f4}). In these three bins, the bottom panels of Fig. \ref{a1f4} show that the DIG spectra (star-like points) have an emission line (blue or red line) not seen in the corresponding typical parasitic spectra (black curve). 

Different opening angles were tested for the pie-shape centred on the parasitic rings, but they always gave the same results. First, that the extracted spectra of the parasitic rings did not vary significantly, which means that the parasitic emission is not just random noise and that a typical spectra can be extracted. Second, when compared to the parasitic spectra, the DIG spectra always showed a $3\sigma$ detection for the same four emission lines mentioned in the previous paragraph. These emission lines lie along the major axis of the galaxy, are not from the parasitic emission, and therefore can be interpreted as emission from the DIG. 

The integrated flux of the first, blue, line ($r=11.5'\pm0.6'$) gives an EM of $\sim0.14$ cm$^{-6}$ pc, and those with the red line ($r=11.5'\pm0.6'$, $12.8'\pm0.7'$, $19.0'\pm0.9'$) give an EM of $\sim0.10$ cm$^{-6}$ pc. Table \ref{a1_t3} shows these results. The EMs were calculated by subtracting the contamination spectra from the DIG profiles, then correcting for the filter dependance and integrating the profiles. The $1\sigma$ upper limit for the EM is $\sim0.03$ cm$^{-6}$ pc. This was calculated by taking the standard deviation ($\sigma$) between the spectrum resulting from the pie-shape centred on the galaxy, and that centred on the parasitic emission, for all the radii bins beyond 11$'$. For the three radii bins where DIG was detected, the emission line profile was omitted before calculating $\sigma$. From the averaged $\sigma$, and a typical width for the extended DIG emission lines, the $1\sigma$ upper limit for the EMs was found to be $\sim$ 0.03 cm$^{-6}$ pc. Hence, the three emission lines seen have EMs on the order of $3\sigma$. Any emission below this limit was not considered as DIG emission. The error of the emission line fluxes is thought to be on the order of $\sigma$, therefore of $\sim30$ per cent. 

\subsection{Nature of the emission lines}

The extended DIG emission line found in the first radii interval ($r\sim11.5'$) and identified with a blue line in Fig 3 should be the H$\alpha$ line at $\lambda\sim6572$ \AA, according to a velocity pattern continuity argument. Due to the periodicity of FP spectrum, in this figure, the wings of this line are seen on both extremity of the spectrum. This is the phase jump effect. Phase jumps are common in FPs when the filter bandpass is larger than the FSR of the FP.

\begin{figure}
\centering
\begin{minipage}[c]{0.99\linewidth}
 \centering \includegraphics[width=\linewidth]{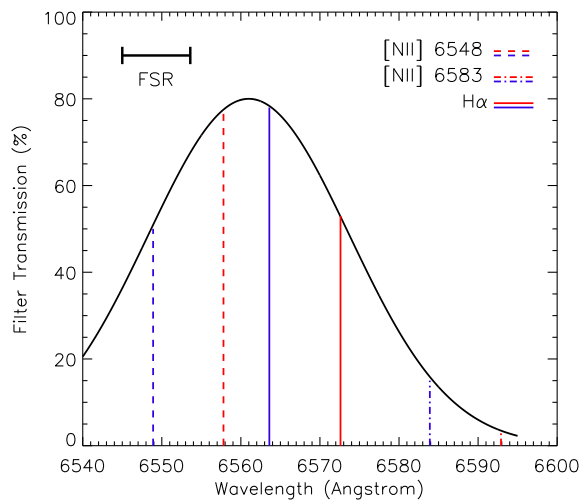}
\end{minipage}
\caption{Interference filter transmission profile (black curve), shown in per cent. The positions of the redshifted (red) and the blueshifted (blue) [N\thinspace{\sc ii}]$\lambda\lambda$ 6548, 6583\AA\ and H$\alpha$ lines are also shown. NGC 253 has a systemic radial velocity of $\sim241$ km s$^{-1}$. The redshifted velocities were calculated using a radial velocity of $\sim450$ km s$^{-1}$ and the blueshifted velocities were calculated using a radial velocity of $\sim40$ km s$^{-1}$. As can be seen, the redshifted [N\thinspace{\sc ii}]$\lambda$ 6548\AA\ is almost perfectly transmitted, much more than the redshifted H$\alpha$ line. The FSR of the FP (8.6 \AA) is also shown in the upper-left corner.}
\label{a1f3add}
\end{figure}

More precisely, a phase jump, also known as interordre confusion of neighbouring FP orders \citep[see][]{Bla199225,Dai2006367}, appears because the interference pattern of the FP will transmit (for a given spacing of the plates) all wavelengths $\lambda$ such that $\lambda=\lambda_\mathrm{0}\pm\rm{n}\times\rm{FSR}$ where $\rm{n}$ is an integer value, providing that these wavelengths are also transmited by the intereference filter used for selecting the wavelength range to be scanned with the FP and to limit the number of phase jumps (see Fig. \ref{a1f3add}). Essentially, the FP acts like a secondary filter, with a transmission peak occurring every 8.6 \AA . In other words, light first travels through the interference filter, which affects the intensity of the light, depending on the transmission probability (Fig. \ref{a1f3add}). Then, light travels through the FP, where phase jumps can occur, but do not affect the intensity of the light. Therefore, the probability of having phase jumps of a specific emission line depends only on whether the probability of it passing through the interference filter is high enough.   

In this study, we have choosen the most adequate interference filter to cover the wavelengh range of the H$\alpha$ line in NGC 253, given its systemic velocity and we study the FP interference order corresponding to the wavelength range between $\sim6564$ \AA\ and $\sim6572$ \AA, covering H$\alpha$ velocities between 40 and 450 km s$^{-1}$ (see Section 2.1). Hence, the spectrum is plotted between $6564$ \AA\ and $6572$ \AA, but there could be emission from neighbooring free spectral ranges: for instance, an emission line detected at $\lambda=6570$ \AA\ could be emitted at $\lambda=6570\pm\rm{n}\times8.6$ \AA.

The DIG emission lines identified by a red line in Fig. \ref{a1f4} at $\lambda\sim6566.4$ \AA\ cannot be from H$\alpha$ since the velocities implied are not consistent with the velocity pattern of the galaxy. They are rather emission from the [N\thinspace{\sc ii}] line ($\lambda$ 6548 \AA ) redshifted, and are probably observed around this wavelength due to a phase jump.

Fig. \ref{a1f3add} shows the filter transmission profile, and the locations of the blueshifted and redshifted emission lines of [N\thinspace{\sc ii}]$\lambda\lambda$ 6548, 6583\AA\ and H$\alpha$ lines. This figure clearly shows that on the receding side of the galaxy (redshifted), the [N\thinspace{\sc ii}]$\lambda$ 6548\AA\ line is more likely to be transmitted than the H$\alpha$ line. The line could be emitted at 6566.4 \AA\ or at 6557.8 \AA\ (in case of a phase jump) or even at 6549.2 \AA\ (in case of multiple phase jumps). 

Due to the interference filter transmission, this line is most likely emitted at 6566.4 \AA\ (~75 per cent) or at 6557.8 \AA\ (78 per cent). In the first case, the line we see at 6566.4 \AA\ is an extremely redshifted [N\thinspace{\sc ii}] line with a radial velocity of $\sim850$ km s$^{-1}$. This radial velocity is not consistent with the galaxy's velocity field, but could be interpreted as an accretion phenomenon. We examine this possibility in Section 7.2.1. In the second case, the velocity ($\sim400$ km s$^{-1}$) is compatible with a declining rotation curve. We use this assumption in Sections 5 and 6. Why no extended DIG emission was seen on the approaching side of the galaxy, as well as why no H$\alpha$ was seen on the receding side beyond $11.5'$ will be looked at in Section 7.2.1.

\section{\large{Kinematical parameters and rotation curve}}\label{a1c4}

In order to derive the rotation curve of NGC 253 from the analysis of the velocity field, the GIPSY task ROTCUR \citep{Beg1989223} was used. The objective is to extract the orientation parameters (dynamical centre ($x_\mathrm{0}$, $y_\mathrm{0}$), systemic velocity $V_\mathrm{sys}$, inclination \textit{i} and position angle PA) that best reproduce the observed velocity field. As mentioned earlier, the velocity field smoothed with a gaussian function of $w_\mathrm{\lambda}=55''$ is used. The parameters are found by a least square method between the observed velocity field and the tilted rings derived by ROTCUR.

There are three main steps with the GIPSY program. Since they are correlated, the dynamical centre and systemic velocity are found first by keeping \textit{i} and PA fixed at their initial values. The initial values are the ones derived by the H\thinspace{\sc i} study of \citet{Puc1991101}. Data in an opening angle of 65$^o$ around the minor axis are excluded in order to minimize errors due to deprojection effects. A cosine-square weighting function is then used on the rest of the data, maximizing the weight of the points around the major axis. From this, the dynamical centre is found to be located at $\alpha=00:47:35.2$ and $\delta=-25:17:07.6$, in the World Coordinate System. A systemic velocity of $241.2\pm5.4$ km s$^{-1}$ is also found. This value is compatible with the values derived from H\thinspace{\sc i} by \citet{Puc1991101} ($\sim245$ km s$^{-1}$) and from H$\alpha$ by \citet{Bla1997490} ($243\pm9$ km s$^{-1}$). 

The second step seeks to find the inclination \textit{i} and the position angle PA, while keeping the dynamical centre and systemic velocity fixed at their extracted values. The initial values for the inclination \textit{i} and position angle PA are again those derived by the H\thinspace{\sc i} study of \citet{Puc1991101}. Separate solutions are obtained for the receding and approaching side. A mean value for the inclination \textit{i} of $76.4\pm3.9^o$ and for the position angle PA of $230.1\pm2.1^o$ are derived between a radius of 1$'$ and 10$'$. ROTCUR could not converge for a small number of radii, and the value of the parameters at these radii were extrapolated with the values of nearby data. The gaussian smoothed velocity field extends to a radius of 10$'$, while the original velocity field extends to 11.5$'$, although the later is very clumpy beyond 10$'$. In order to derive the rotation curve from the original velocity field using the derived kinematical parameters, extrapolated values of the parameters were added for radii between 10$'$ and 11.5$'$. The top figures of Fig. \ref{a1f5} show the adopted values for the inclination \textit{i} and position angle PA. The results are consistent with the parameters derived by the H\thinspace{\sc i} study of \citet{Puc1991101}, and also suggest the presence of a warp in the outer regions as seen through the decrease of the position angle at $r\sim5'$. In the inner 5$'$ region, a variation of the position angle is also seen. However, this can be attributed to the perturbations of the nuclear region since NGC 253 is known to be a starburst galaxy. \citet{Puc1991101} find for the H\thinspace{\sc i} data a mean inclination \textit{i} of $\sim72^o$ and a mean position angle PA of $\sim229^o$. 
\begin{figure}
\centering
\begin{minipage}[c]{0.99\linewidth}
 \centering \includegraphics[width=\linewidth]{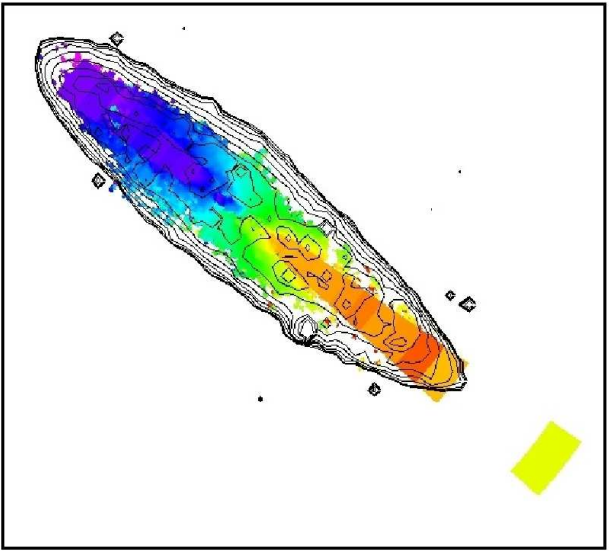}
\end{minipage}
\caption{Velocity field of NGC 253, in which the pie shape used to identify the diffuse emission is shown. Contours of the near-UV GALEX image in a logarithm scale are shown. The outermost contours was taken as that with 0.005 counts/second/pixel.}
\label{a1f4add}
\end{figure}
From the adopted values of the inclination \textit{i} and position angle PA varying with radius, the rotation curve of NGC 253 could be extracted from the original velocity field (without the gaussian smoothing). The warp seen in NGC 253 \citep{Puc1991101} can be accounted for by letting the inclination and the position angle vary with radius. The bottom-left image of Fig. \ref{a1f5} illustrates the rotation curve derived, along with the solutions found for the approaching and receding side. Finally, the bottom-right image gives the adopted rotation curve of the galaxy, with the proper error bars. The error bars correspond to the biggest difference between the rotation curve of the whole galaxy and that of one of the sides, or, if larger, the intrinsic error determined by the ROTCUR task. 
 
The velocities determined for the DIG component from the previous section are shown in the bottom-right panel of Fig. \ref{a1f5} (diamonds). The kinematical parameters used to derive the rotation velocity from the line of sight velocity were those at the maximum radius of 690$''$. The error bars in the abscissae direction were chosen as the half width of the radii interval from which the emission line was extracted, while the one in the vertical direction was carefully chosen as the one resulting from the dispersion of the emission line since the largest uncertainty is on the precise location of the profile, rather than on the variation in the kinematical parameters. 

In order to derive the velocities of the DIG, we assume that circular motions remain at large radii. Non-circular motions have been studied in the central regions of galaxies, where phenomena such as bars have been found, but it remains difficult to study these motions in the outskirts of galaxies. Deep surveys of matter at large radii are needed. The study by \citet{Dic2008135} found an extended H$\alpha$ rotation curve for NGC 7793, another galaxy of the Sculptor group. The H$\alpha$ emission extended out to the H\thinspace{\sc i}, and showed a rotation curve consistent with that of the H\thinspace{\sc i}. When deriving the kinematical parameters of NGC 253, no significant non-circular motions were seen at $r\ge7'$. For NGC 253, we examined the surface-brightness isophotes of the near and far ultra-violet (UV) maps, taken from the Galaxy Evolution Explorer (GALEX) survey. On the receding side of the galaxy, the isophotes extended out to a radius of about 13$'$. This wavelength traces hot stars, which could be a source of ionization and could explain our extended observations. The isophotes showed distortions in the inner regions, but were symmetric out to 13$'$ (see the near-UV contours shown in Fig. \ref{a1f4add}). This suggest, that at least for the two inner bins where we detected diffuse gas ($r=11.5'$, 12.8$'$), we can treat the rotation as been circular. For the outermost point ($r=19.0'$), our observations do not allow us to make a definitive answer. 

A deep optical image of NGC 253 was taken by \citet[][]{Mal199714}, and is presented in Fig. 2 of \citet{Boo2005431}. This image supports the idea that the stellar disc of NGC 253 extends well beyond the H\thinspace{\sc i} disc, but the image also suggests that there are some disturbances on the south-west side of the galaxy. This could contradict the results regarding the Galex isophotes. However, we prefer to keep with the assumption of circular orbits as in \citet[][]{Bla1997490}'s paper, and see what kind of results we obtain. 

\begin{figure*}
\centering
\begin{minipage}[c]{0.47\linewidth}
 \centering \includegraphics[width=\linewidth]{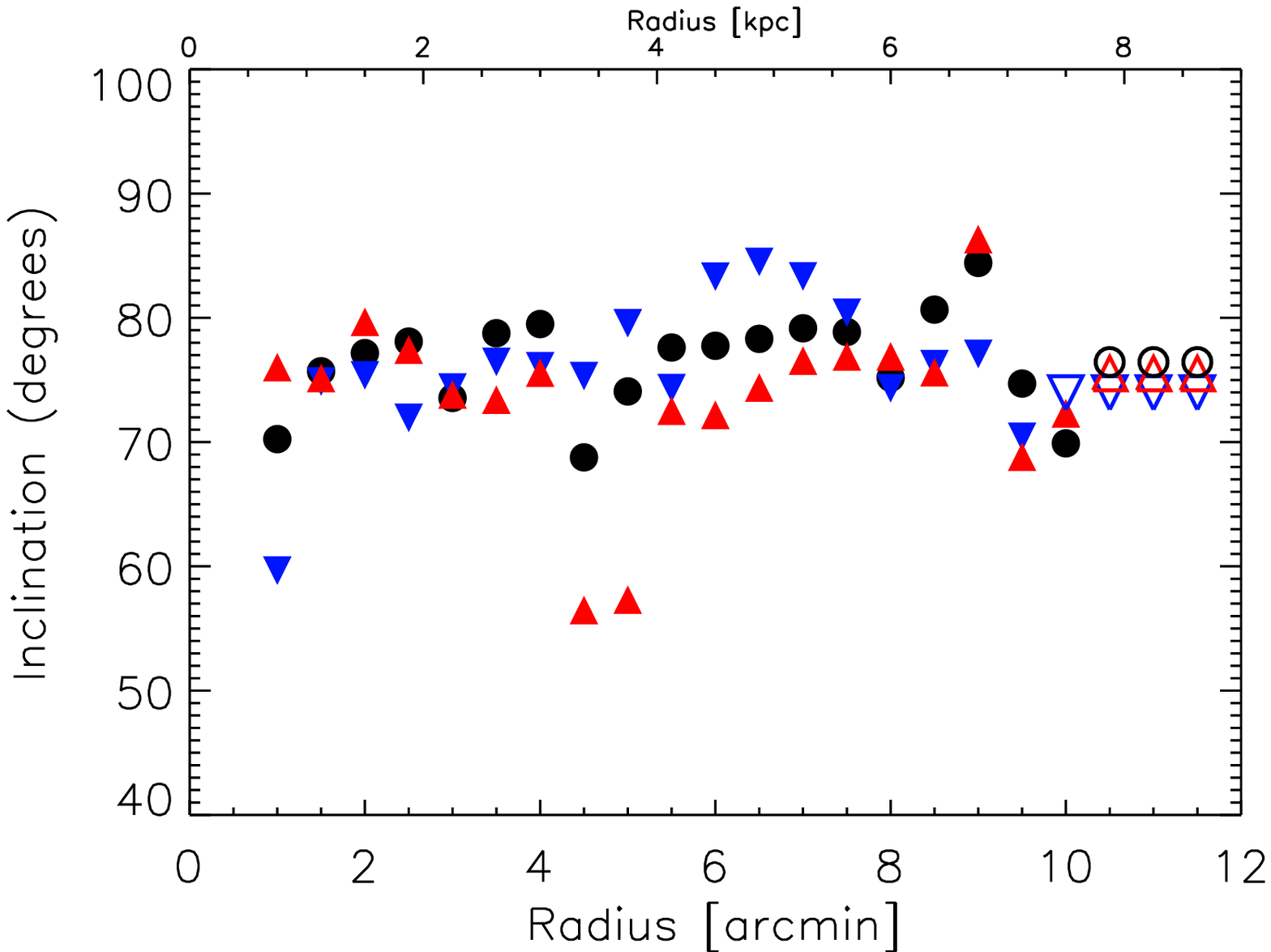}
\end{minipage}
\begin{minipage}[c]{0.47\linewidth}
 \centering \includegraphics[width=\linewidth]{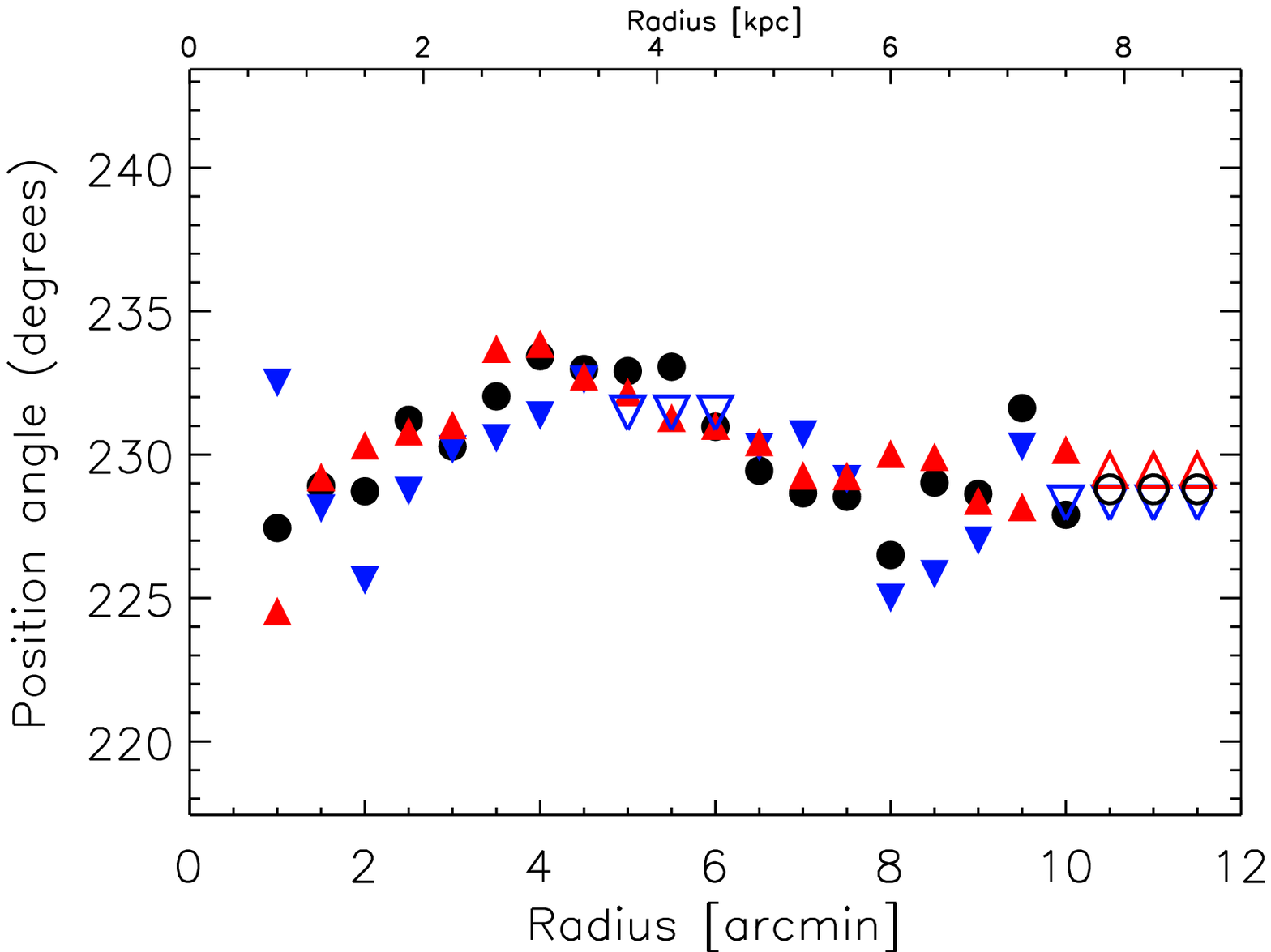}
\end{minipage}
\begin{minipage}[c]{0.47\linewidth}
 \centering \includegraphics[width=\linewidth]{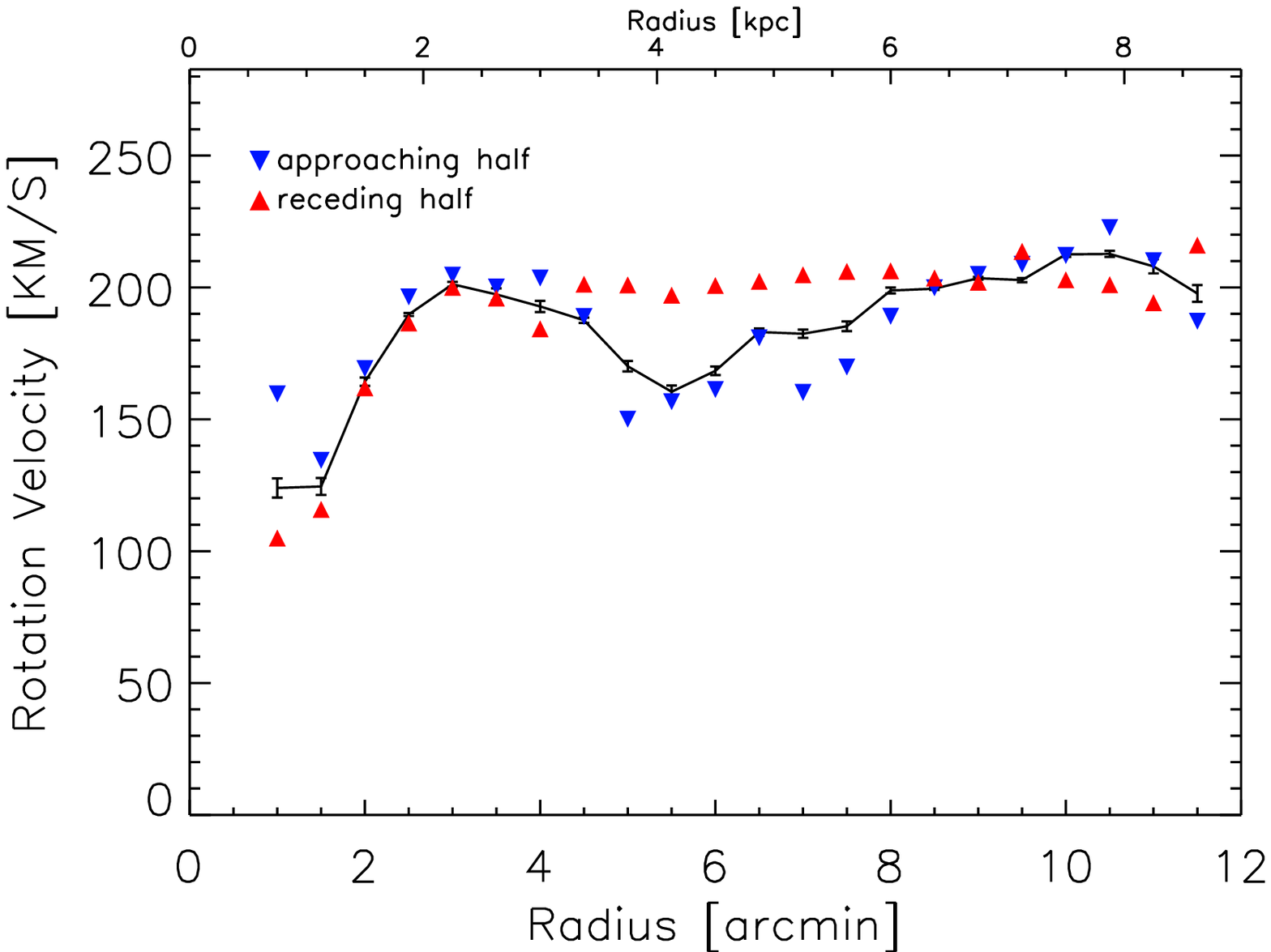}
\end{minipage}
\begin{minipage}[c]{0.47\linewidth}
 \centering \includegraphics[width=\linewidth]{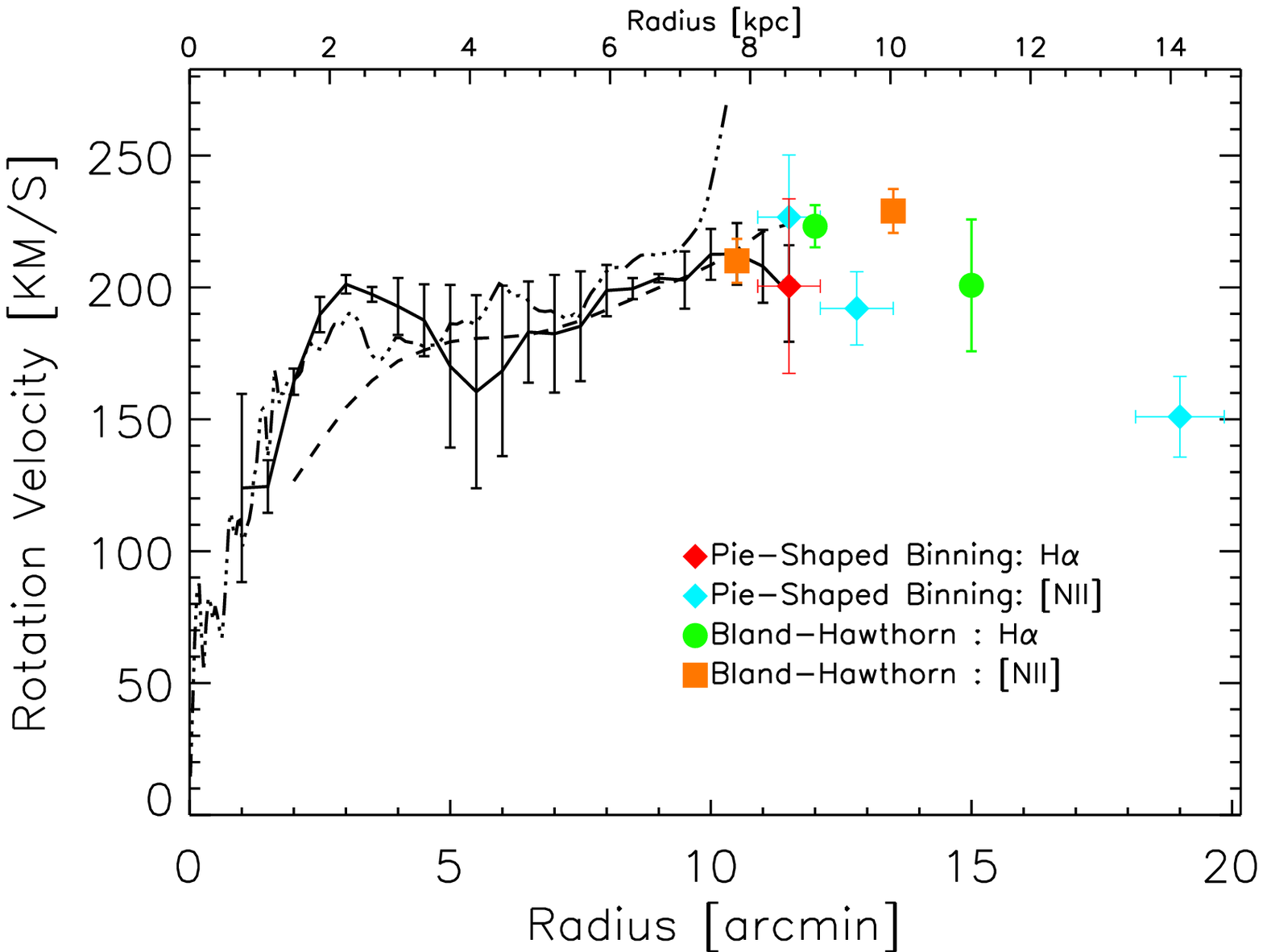}
\end{minipage}
\caption{Top: Kinematical parameters of NGC 253 derived from the gaussian smoothed ($w_\mathrm{\lambda}=55''$) velocity field for both sides (circles) and separately for the receding (red triangles) and for the approaching (blue triangles) sides. The non-filled symbols represent the extrapolated parameters, see Section 4 for more details. Bottom-left: Rotation curve derived using the kinematical parameters of the top figures as a function of radius, with a systemic velocity of $V_\mathrm{sys}=241.2$ km s$^{-1}$, and the non gaussian smoothed velocity field. The rotation curve obtained using both sides at once is represented by the continuous line, along with the derived error bars, while the receding (red) and approaching (blue) sides are represented by the triangles. Bottom-right: Final adopted rotation curve, with the proper error bars as explained in the text (continuous line). The slashed line represents the H\thinspace{\sc i} data from \citet{Puc1991101}, and the double-dot slashed line represents the H$\alpha$ data from \citet{Arn1995110}. The colored points illustrate the results of \citet{Bla1997490} for [N\thinspace{\sc ii}] (orange squares) and for H$\alpha$ (green circles) beyond the H\thinspace{\sc i} data, as well as those derived from this study (blue and red diamonds).}
\label{a1f5}
\end{figure*}
The adopted values of the rotation velocity and the associated errors are given in Table \ref{a1_t2} and Table \ref{a1_t3}. In order to compare our results, with the previous ones, the H\thinspace{\sc i} rotation curve found by \citet{Puc1991101} is illustrated in Fig. \ref{a1f5}, as well as the extended data points found by \citet{Bla1997490}, where the inclination and systemic velocity of this study were used. For the later, see Table 3. Finally, the H$\alpha$ rotation curve derived by \citet{Arn1995110}, using long-slit observations, is also illustrated in the figure. The latter study did not correct the rotation curve for the inclination or the PA, and therefore did not consider the variation in the kinematical parameters with radius. Deviations are therefore expected. Their curve, plotted in Fig. \ref{a1f5}, has been corrected for inclination using the mean value derived for our study. 
\begin{table}
\centering
\caption{Adopted H$\alpha$ rotation curve \label{a1_t2}}
\resizebox{0.46\textwidth}{!}{
\begin{tabular}{@{}cccccc@{}}
\hline
\hline
& & & & & \\
\textit{radius} & \textit{V}$_\mathrm{rot}$ & $\Delta$\textit{V}$_\mathrm{rot}$ & \textit{radius} & \textit{V}$_\mathrm{rot}$ & $\Delta$\textit{V}$_\mathrm{rot}$ \\
(arcsec) & (km s$^{-1}$) & (km s$^{-1}$) & (arcsec) & (km s$^{-1}$) & (km s$^{-1}$) \\
\hline
& & & & & \\
60 & 123.96 & 35.66 & 390 & 183.09 & 19.21 \\
90 & 124.54 & 9.97 & 420 & 182.45 & 22.31 \\
120 & 164.25 & 4.97 & 450 & 185.27 & 20.80 \\
150 & 189.73 & 6.71 & 480 & 198.82 & 9.74 \\
180 & 201.20 & 3.50 & 510 & 199.55 & 4.01 \\
210 & 197.36 & 2.82 & 540 & 203.52 & 1.54 \\
240 & 192.80 & 10.79 & 570 & 202.81 & 10.88 \\
270 & 187.55 & 13.65 & 600 & 212.54 & 9.65 \\
300 & 170.12 & 30.85 & 630 & 212.72 & 11.67 \\
330 & 160.46 & 36.60 & 660 & 208.00 & 13.81 \\
360 & 168.38 & 32.31 & 690 & 197.72 & 18.30 \\
\hline
\end{tabular}}
\end{table}
\begin{table}
\centering
\caption{Derived DIG rotation velocities and those of Bland-Hawthorn et al. (1997)} \label{a1_t3}
\resizebox{0.46\textwidth}{!}{
\begin{tabular}{@{}ccccc@{}}
\hline
\hline
This Study & & &\\
\textit{radius} & $\Delta$\textit{r} & \textit{V}$_\mathrm{rot}$ & $\Delta$\textit{V}$_\mathrm{rot}$ & EM \\
(arcsec) & (arcsec) & (km s$^{-1}$) & (km s$^{-1}$)  & cm$^{-6}$ pc\\
\hline
& & & \\
690$^a$ & 36 & 200.51 & 33.10 & 0.14 \\
690$^b$ & 36 & 226.64 & 23.60 & 0.10 \\
768$^b$ & 42 & 187.06 & 13.90 & 0.10 \\
1140$^b$ & 51 & 150.98 & 15.25 & 0.10 \\
\hline
& & & \\
BH 1997 & & & \\
630$^b$ & 30 & 210.1 & 8.3 & 0.18 \\
720$^a$ & 25 & 223.2 & 8.0 & 0.16 \\
810$^b$ & 20 & 229.0 & 8.3 & 0.09 \\
900$^a$ & 15 & 200.8 & 25.0 & 0.08 \\
\hline
$^a$ H$\alpha$ & & & \\
$^b$ [N\thinspace{\sc ii}] & & & \\
\end{tabular}
}
\end{table}

\section{\large{Mass models}}\label{a1c5}

In order to determine the impact of the DIG component velocity points at large radii, two mass models analysis have been obtained for NGC 253. The first has been performed on the data taken without the three points of the DIG while the second included those extended velocities. Furthermore, in order to compare our results with those found for the H\thinspace{\sc i} data by \citet{Puc1991101}, the analysis of the mass distribution must be the same. Hence, another mass model analysis has been obtained using the data taken from \citet{Puc1991101}. A dark halo represented by an isothermal sphere is used. The density of the isothermal sphere has a flat slope in the central regions. Observationally, the data favour models with flat slopes rather than those with \textit{cuspy} cores such as the Navarro, Frenk \& White (NFW) profile \citep[e.g.][]{Bla2001121,deB2002385,deB2003340,Kas2006162}.

Three parameters must be specified: two for the dark halo, namely a one-dimentional velocity dispersion $\sigma$ and a core radius ${r_\mathrm{c}}$, related together by the central density $\rho_\mathrm{0}=9\sigma^2/4{\pi}G{r_\mathrm{c}}^2$, and the mass-to-light ratio (\textit{M}/\textit{L})$_\star$ of the stellar disc. The papers by \citet{Car1985294} and \citet{Car1985299}, show in greater detail the method used. A least square fit is applied to determine the three parameters. In order to compare with the H\thinspace{\sc i} study, the stellar disc is described by the surface brightness profile in the \textit{B} band, as in \citet{Puc1991101}. The gaseous component is also taken as the one derived in the H\thinspace{\sc i} study. The latter includes the 4/3 multiplication factor to account for the H$_\mathrm{e}$ contribution. 

The mass model is designed to put higher weight on the data points that have small errors. In the case of the H\thinspace{\sc i}, the data at intermediate radii, between 3 and 5.5 kpc, have much smaller errors than the central and outer points. On the other hand, in the case of the H$\alpha$ data, the mass model tended to fit as best as possible the central areas which had the smallest errors. This yielded very different results. Hence, in order to better compare the H\thinspace{\sc i} and the H$\alpha$ data, \emph{equal weight was given to all data points}. 

\begin{figure}
\centering
\begin{minipage}[c]{0.99\linewidth}
 \centering \includegraphics[width=\linewidth]{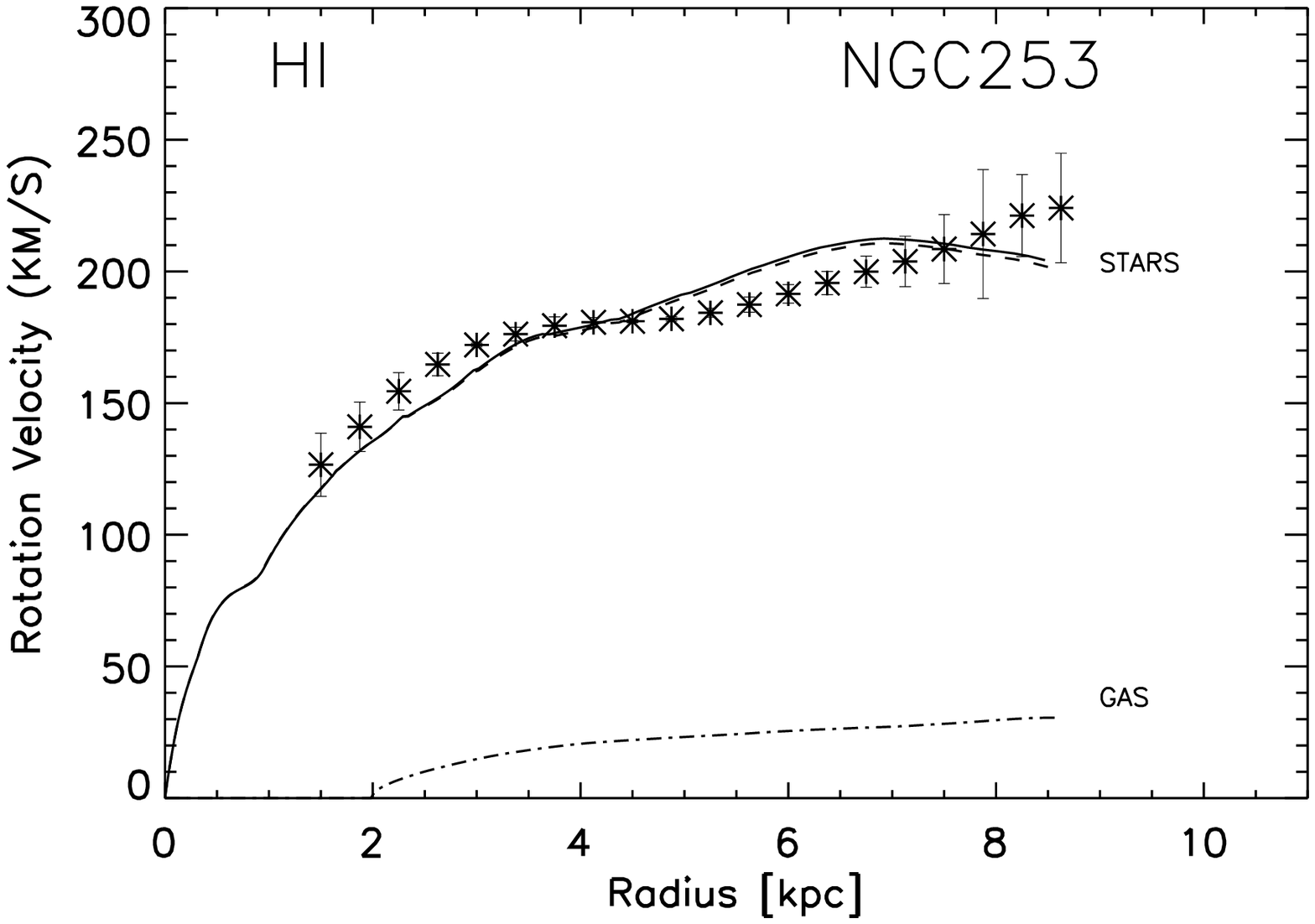}
\end{minipage}
\begin{minipage}[c]{0.99\linewidth}
 \centering \includegraphics[width=\linewidth]{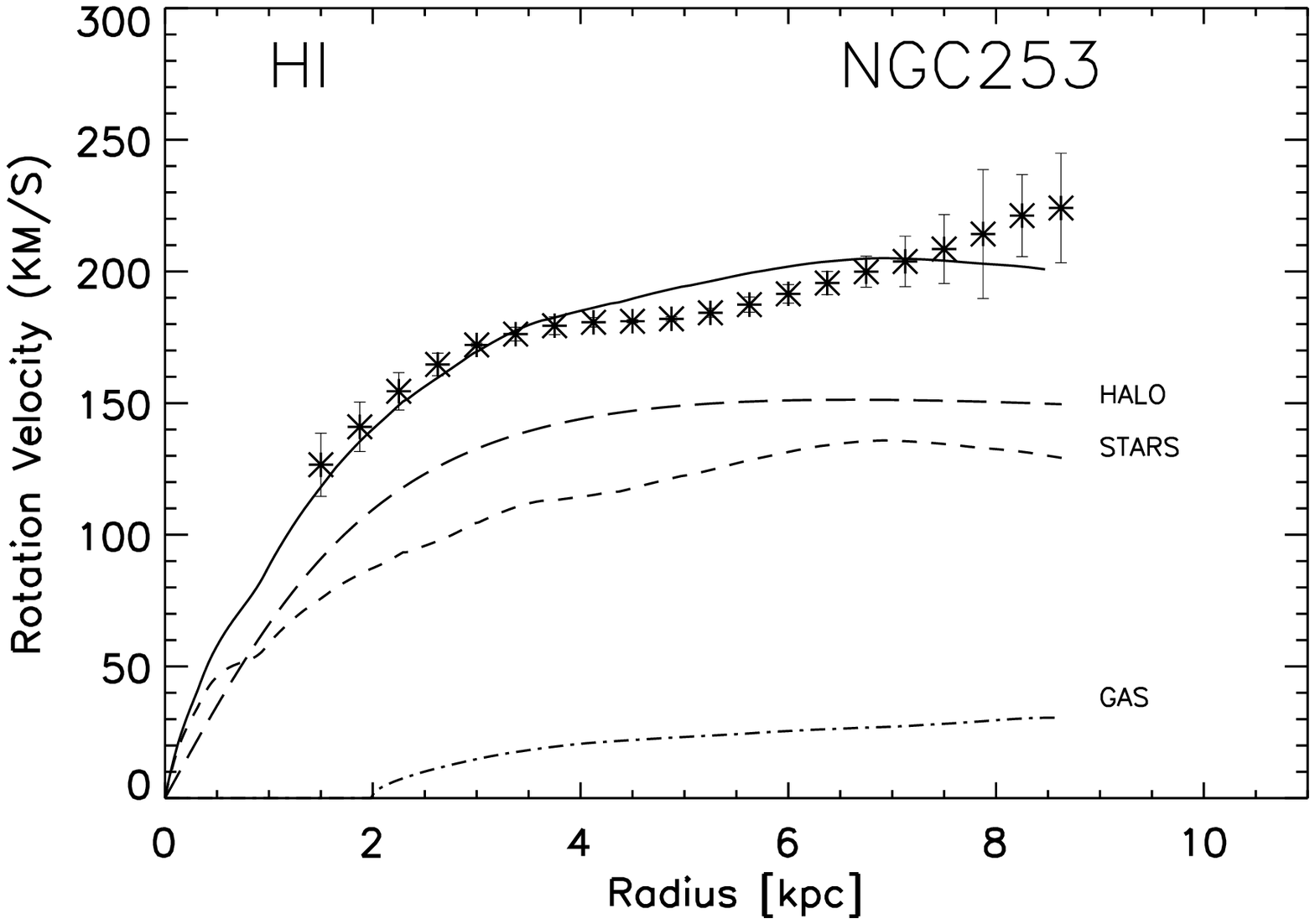}
\end{minipage}
\caption{Mass model analysis for the adopted H\thinspace{\sc i} rotation curve (star-like points) of NGC 253. The data were taken from \citet{Puc1991101}. An isothermal sphere was used for the dark halo. A first analysis was performed using a maximum stellar disc (top), and a second using a sub-maximum stellar disc (bottom). The stellar disc is plotted with a small-dashed line, the gaseous component with a dot-dashed line, the halo with a long-dashed line and the total adjusted rotation curve with the continuous line. The parameters extracted are in Table. \ref{a1_t4}.}
\label{a1f6a}
\end{figure}

\begin{figure}
\centering
\begin{minipage}[c]{0.99\linewidth}
 \centering \includegraphics[width=\linewidth]{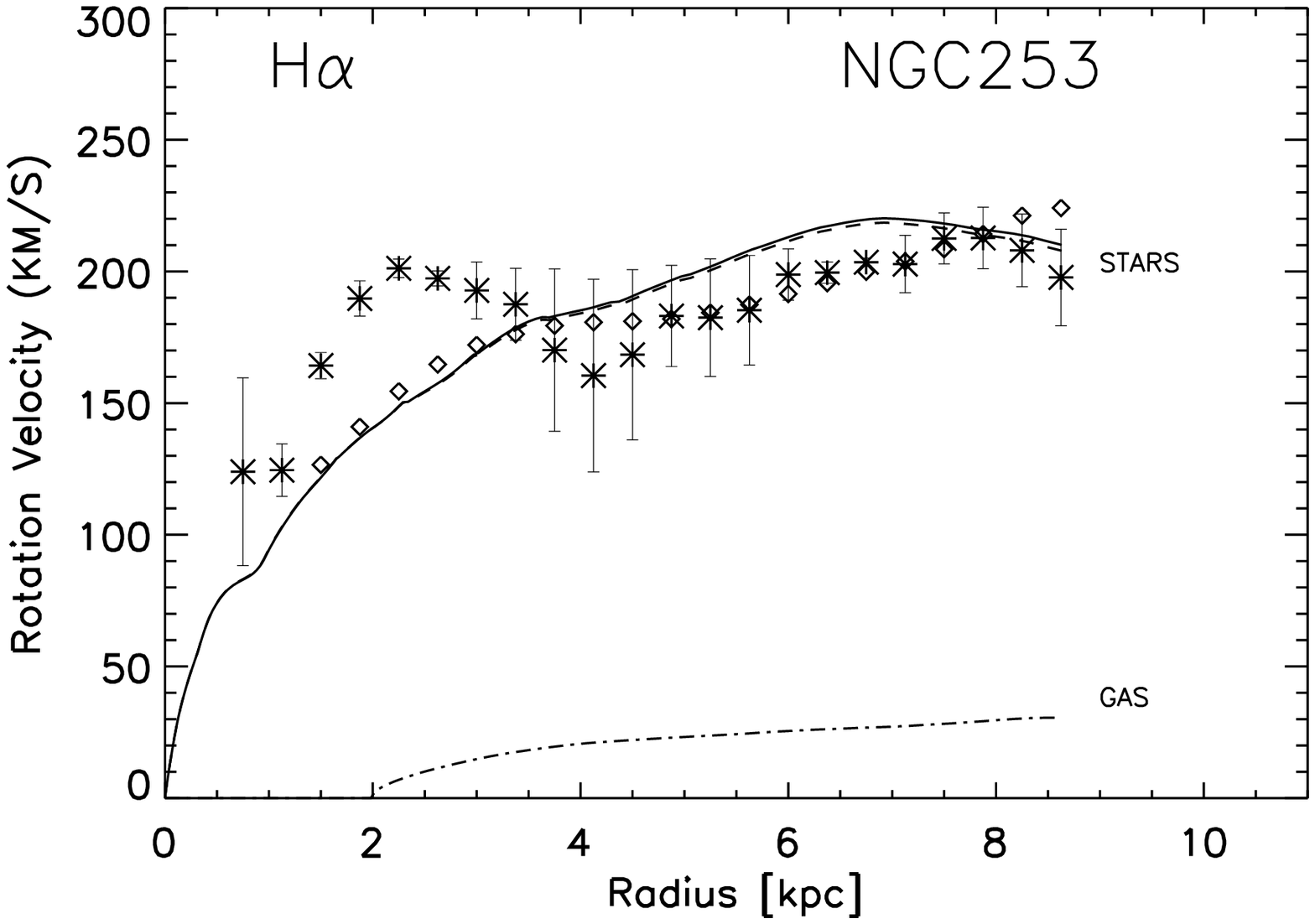}
\end{minipage}
\begin{minipage}[c]{0.99\linewidth}
 \centering \includegraphics[width=\linewidth]{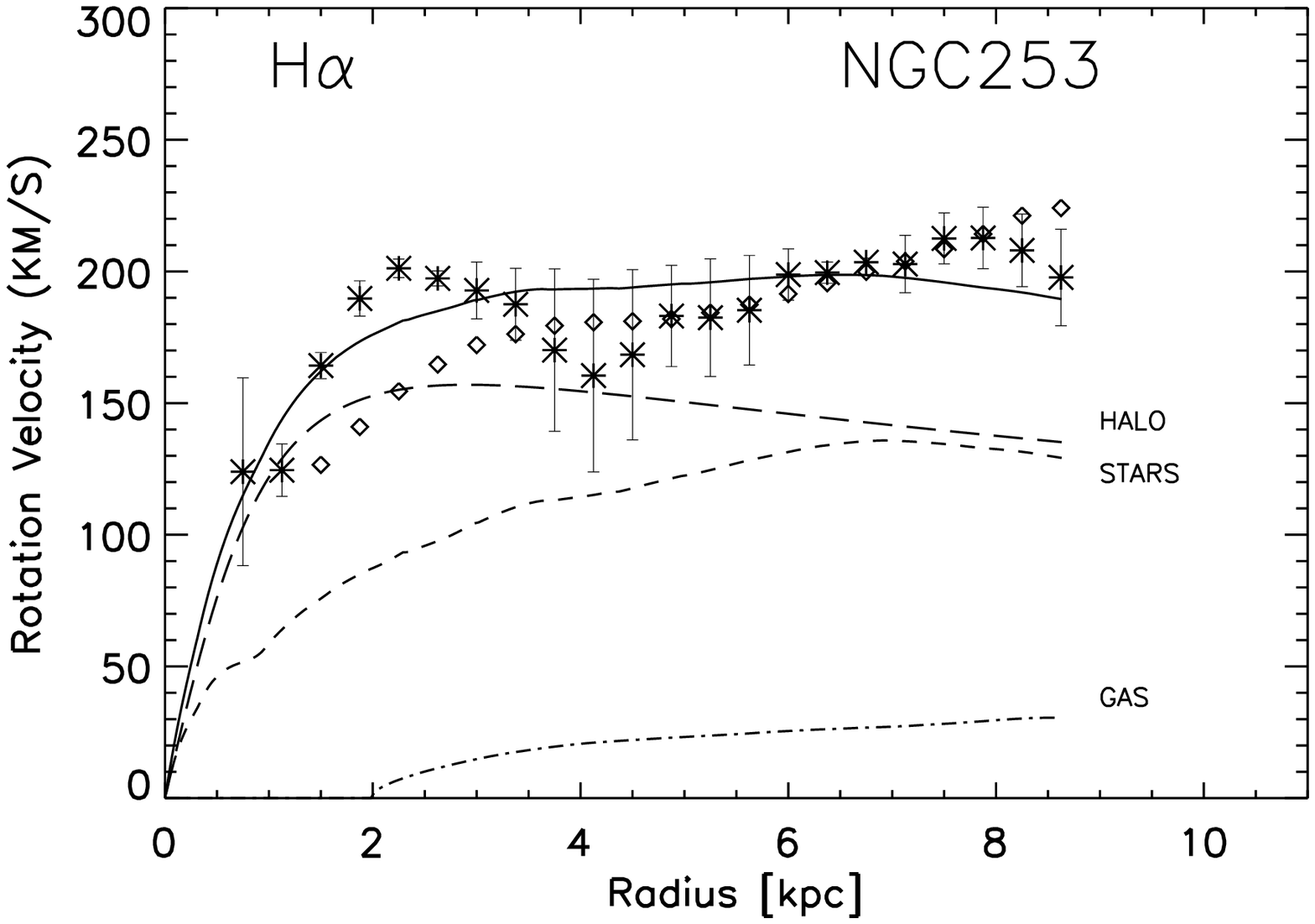}
\end{minipage}
\caption{Mass model analysis for the adopted H$\alpha$ rotation curve from this study (star-like points) of NGC 253, without the DIG [N\thinspace{\sc ii}] points. Same caption as Fig. \ref{a1f6a} with a maximum stellar disc (top) and a sub-maximum stellar disc (bottom). The H\thinspace{\sc i} data from \citet{Puc1991101} are also plotted with the diamond-shaped points.}
\label{a1f6b}
\end{figure}

\begin{figure}
\centering
\begin{minipage}[c]{0.99\linewidth}
 \centering \includegraphics[width=\linewidth]{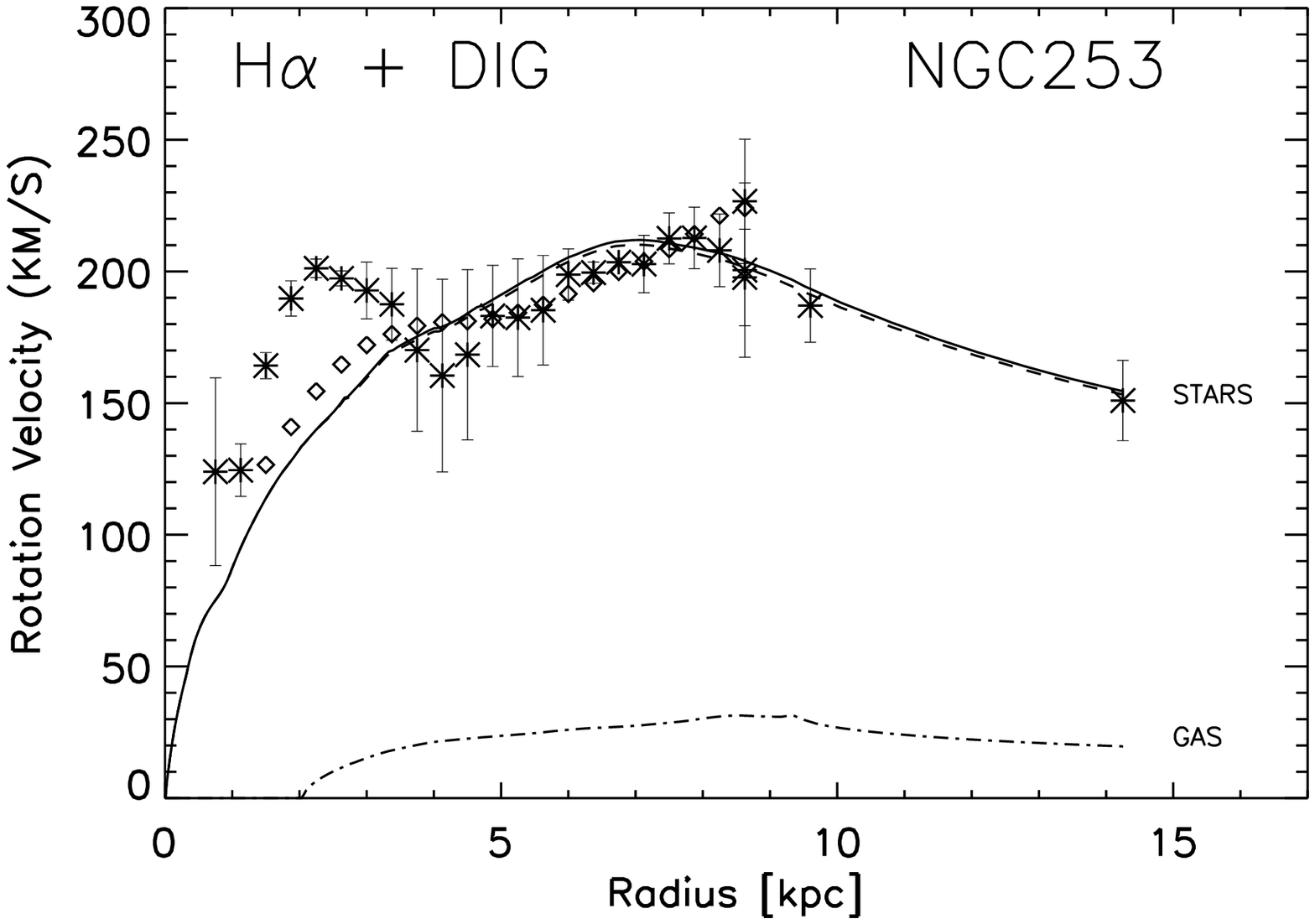}
\end{minipage}
\begin{minipage}[c]{0.99\linewidth}
 \centering \includegraphics[width=\linewidth]{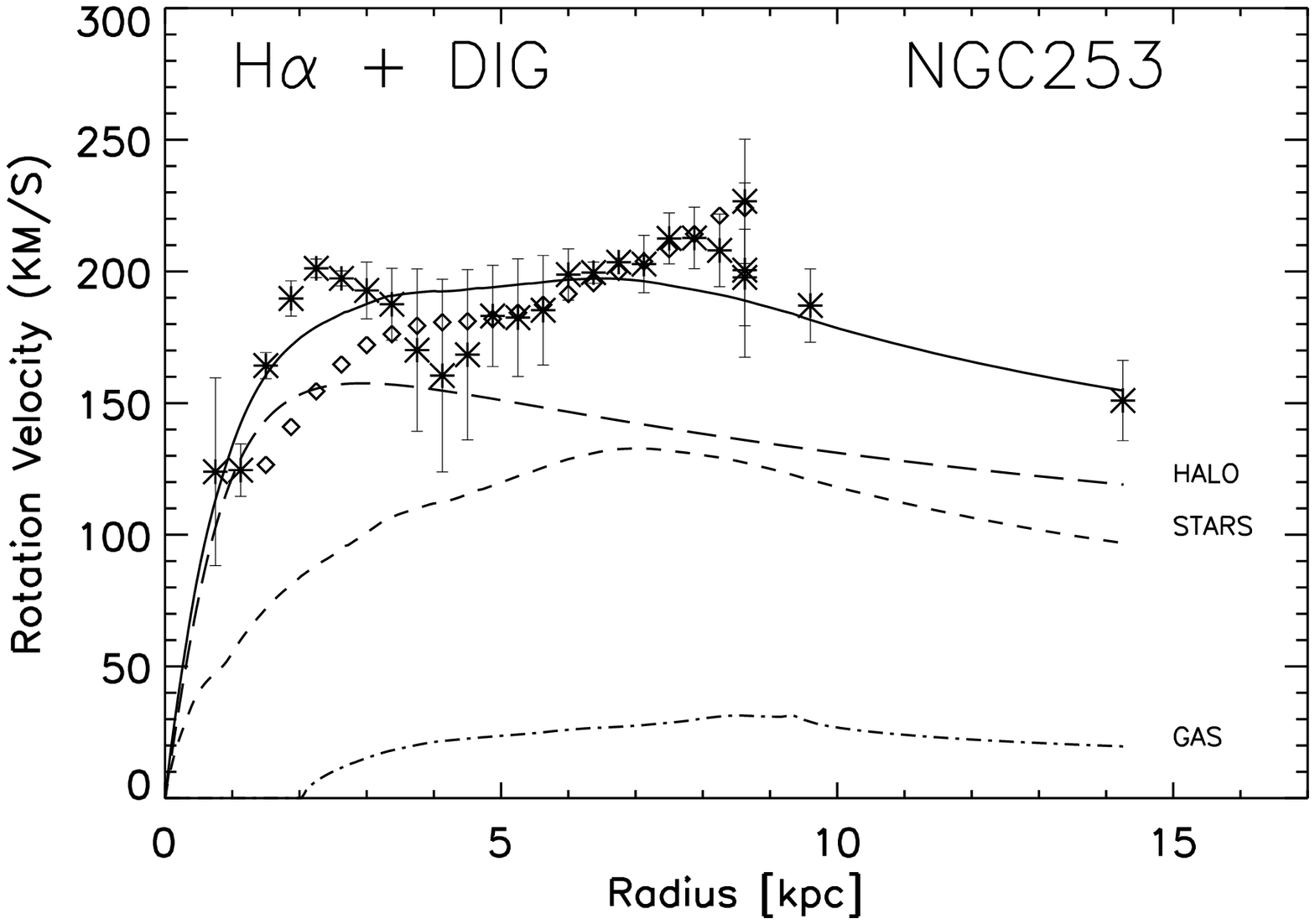}
\end{minipage}
\caption{Mass model analysis for the adopted H$\alpha$ rotation curve with the DIG [N\thinspace{\sc ii}] + H$\alpha$ points (star-like points) of NGC 253. Same caption as Fig. \ref{a1f6a} with a maximum stellar disc (top) and a sub-maximum stellar disc (bottom). The H\thinspace{\sc i} data from \citet{Puc1991101} are also plotted with the diamond-shaped points.}
\label{a1f6c}
\end{figure}

First, a maximum disc model with no halo component was obtained (see top of Fig. \ref{a1f6a}, Fig. \ref{a1f6b}, Fig. \ref{a1f6c} and Table \ref{a1_t4}). Second, a sub-maximum disc was obtained. Here, (\textit{M}/\textit{L})$_\star$ is fixed to the value that allows the maximum velocity of the stellar disc to be 0.6 times the maximum velocity of the rotation curve (H\thinspace{\sc i} or H$\alpha$) as suggested by the study of \citet{Cou1999513}. Indeed, the (\textit{M}/\textit{L})$_\star$ determined from the maximum disc is usually too high to agree with the predictions of the stellar populations models \citep{Bel2001550}. See bottom of Fig. \ref{a1f6a}, Fig. \ref{a1f6b}, Fig. \ref{a1f6c} and Table \ref{a1_t4}.
\begin{table}
\centering

\caption{Mass models for NGC 253}\label{a1_t4}
\begin{minipage}[t]{14cm}
    \renewcommand{\footnoterule}{}
\begin{tabular}{@{}llll@{}}
\hline
\hline
& H\thinspace{\sc i} RC\footnote[1]{Puche et al. (1991)} & 36-cm H$\alpha$ RC& 36-cm H$\alpha$ RC\\
& & without DIG & with DIG \\
\hline
\textbf{Max Disc} & & & \\
& & & \\
(\textit{M}/\textit{L}$_\mathrm{B}$)$\star$\footnote[2]{in units of M$_\odot$/L$_\odot$} & 12.0 & 12.9 & 12.5 \\
& & & \\
Dark halo\footnote[3]{\textit{r}$_\mathrm{c}$ in kpc, $\sigma$ in km s$^{-1}$ and $\rho$$_\mathrm{0}$ in M$_\odot$ pc$^{-3}$} & none & none & none \\
& & & \\
\hline 
\textbf{Sub-Max Disc} & & & \\
& & & \\
(\textit{M}/\textit{L}$_\mathrm{B}$)$\star$ & 5.0 & 5.0 & 5.0 \\
& & & \\
Dark halo & \textit{r}$_\mathrm{c}$ = 2.3 & \textit{r}$_\mathrm{c}$ = 1.0 & \textit{r}$_\mathrm{c}$ = 1.0 \\ 
& $\sigma$ = 2.72 & $\sigma$ =  3.04 & $\sigma$ = 2.99 \\
& $\rho$$_\mathrm{0}$ = 0.286 & $\rho$$_\mathrm{0}$ = 1.579 & $\rho$$_\mathrm{0}$ = 1.564 \\
& & & \\
\hline
\end{tabular}
\end{minipage}
\end{table}

\section{\large{Analysis of the results}}\label{a1c6}
 																					
\subsection{\normalsize Nuclear region}\label{a1c61}

NGC 253 is a starburst spiral galaxy. The central regions have been thoroughly studied. Some suggest that NGC 253 is also an AGN \citep{Pta1997113} and has extensive galactic winds \citep*[see a review by][]{Vei200543}. The paper by Heesen et al. (2009)\nocite{Hee2009494a} also finds that the poloidal component of the magnetic field shows a prominent X-shaped structure centred on the nucleus. \citet{Boo2005431} also find extra-planar H\thinspace{\sc i} on the north-east half of the galaxy, in the shape of a half-ring structure and plumes, with a smaller rotation velocity than the disc. It is suggested that the origin of the extra-planar H\thinspace{\sc i} is related to the starburst activity and the active star formation in the disc. 

The present study finds high H$\alpha$ emission and high velocity dispersions in the nuclear region, which are expected from the strong nuclear activity. Higher-resolution data have also seen a large dispersion in the velocity field \citep[see][]{Zha2001240}, which suggest that the one found in this study is not due to beam smearing. A small bar-like structure is also seen in the velocity field, lagging in velocity by at least 50 km s$^{-1}$ (see top-left image of Fig. \ref{a1f3}; the bar-like structure in the nuclear region appears more blue-shifted than its surroundings). The structure also seems slightly inclined with respect to the major axis. The dimensions are approximately 24$''$ by 88$''$, i.e. 0.3 kpc by 1.1 kpc. The strong activity in the nuclear region of the galaxy could account for these observations, including the presence of a possible bar \citep{Pag2004611} or a possible counter-rotating circumnuclear disc \citep{Zha2001240}. An inclined counter-rotating disc could be overlaying the bulk circular rotating disc and account for the lagging velocities and strong dispersion. \citet*{Rub1992394} and \citet{Rix1992400} have shown for the early-type galaxy NGC 4550, that a natural explanation for a decrease in velocities and increase in dispersion could be the presence of two counter-rotating stellar components, as indicated by the observed double-peaked line profiles. Although NGC 253 is not an early-type galaxy, but a late-type, this explanation could also be appropriate for its gaseous component. 

In Fig. \ref{a1f5}, the H$\alpha$ rotation curve derived from this study is seen to agree well with the H\thinspace{\sc i} rotation curve derived by \citet{Puc1991101}, except for the inner regions. Here, the H$\alpha$ slope is much steeper. This can be understood since the H\thinspace{\sc i} data is of much lower resolution ($\sim68''$) compared to the H$\alpha$ data ($\sim2.8''$), and beam smearing is expected which causes the inner slope to be less steep. Hence, an important advantage is that H$\alpha$ data probes more accurately the inner regions of the rotation curve. 

\subsection{\normalsize Diffuse Ionized Gas}\label{a1c62}

\subsubsection{\normalsize [N\thinspace{\sc ii}] and H$\alpha$ emission}

This study succeeded in detecting the [N\thinspace{\sc ii}] ($\lambda$ 6548 \AA ) emission on the receding side of NGC 253 at radii of 11.5$'$, 12.8$'$ and 19.0$'$, as well as H$\alpha$ emission on the receding side at a radius of 11.5$'$. The details concerning the identification of extended DIG emission are decribed in Section 4.  

No [N\thinspace{\sc ii}] emission was seen on the approaching side. The $3\sigma$ upper limit for the EM is $\sim0.09$ cm$^{-6}$ pc (see Section 4). The presence of the [N\thinspace{\sc ii}] emission on the receding side could be the result of a phase jump in the emission line, and the detection was just above a $3\sigma$ detection. Fig. \ref{a1f3add} shows that the filter transmission on the receding side was almost maximal for the [N\thinspace{\sc ii}] line, about 80 per cent. However, on the approaching side, the transmission of this line drops down to 50 per cent, which would make the $3\sigma$ detection difficult. This could explain why no emission of this line was seen on the approaching side. 

For the receding side, no [N\thinspace{\sc ii}] emission was seen between $12.8'\pm0.7'$ and $19.0'\pm0.9'$. No significant $3\sigma$ nor even $2\sigma$ detection was seen. Between these radii, the parasitic emission varied more, i.e. was noisier. Since the [N\thinspace{\sc ii}] emission that was detected ($r=11.5'$, 12.8$'$, 19.0$'$) was already difficult to detect, being just barely above a $3\sigma$ level, the noisier parasitic emission could explain why no [N\thinspace{\sc ii}] was detected between 12.8$'$ and 19.0$'$. Another explanation could be that the DIG is in the form of discrete clouds, which would mean that emission from it can only be seen at discrete radii. Even if these clouds are embedded in a diffuse background, the EMs from this background might be too low to be detected. 

A question remains unanswered. It concerns the H$\alpha$ emission. In the pie-shape procedure, H$\alpha$ was detected up to a maximum radius of $11.5'\pm0.6'$, which coincides with the maximum radius of the H\thinspace{\sc i} study of \citet{Puc1991101}. The strength of the diffuse emission line in this radii bin (receding side, $r=11.5'$) is about a factor of 10 lower than the strength of the line in the previous radii bins. No H$\alpha$ emission is observed beyond, on both sides of the galaxy. Fig. \ref{a1f3add} shows that for H$\alpha$, the transmission probability is only about 50 per cent for the receding side, and about 80 per cent for the approaching side. Hence, detecting H$\alpha$ emission beyond the H\thinspace{\sc i} is much more probable on the approaching side with the observing tools of this study, and yet, no $3\sigma$ detection was seen. 

Bland-Hawthorn et al. (1997) mention the possibility that their detections on the receding side could be explained by accretion of a faint gas-rich object at large radius. This is supported by an image only showing extended stellar light at large radii on the receding side (see Bland-Hawthorn et al. 1997, for more details). This could explain why no extended DIG was seen on the approaching side.

Considering that we were only able to detect emission at specific radii and not continuously across all radii, the clumpiness of the gas would argue for an accretion phenomenon. It has been suggested that the extended disc in M31 is the result of accretion \citep{Iba2005634,Pen2006650}. Also, as mentioned in Section 4.3., the [N\thinspace{\sc ii}] line we see at 6566.4 \AA\ and  could be an extremely redshifted line with a radial velocity of $\sim850$ km s$^{-1}$, which would again argue for accretion of a gas rich object onto the galaxy. 

However, our observations do not entirely support this idea. First, the fact that emission was only seen along the major axis argues that this emission could be associated with the galaxy. Second, as mentioned earlier, the isophotes in the UV GALEX images are symmetric out to $r=13'$. If there was an accretion phenomenon, the isophotes would have been disrupted. This is supported by the isophotes in the $B$ band image taken from the Digitized Sky Surveys. Images of this survey were taken from the Oschin Schmidt Telescope on Polamar Mountain and the UK Schmidt Telescope. These isophotes also extend until $r=13'$ and remain symmetric in the outer radii. Third, our H$\alpha$ rotation curve is consistent with the H\thinspace{\sc i} rotation curve. Both curves seem to behave normally at extended radii, supporting the idea that their are no major disruptions. Fourth, the rotation velocity we obtain for the diffuse H$\alpha$ emission at $r=11.5$ is consistent with that of the phase jumped [N\thinspace{\sc ii}] emission at the same radius. The [N\thinspace{\sc ii}] emission therefore seems to trace the same rotation as H$\alpha$. If the H$\alpha$ is not disrupted up until 11.5$'$, then the [N\thinspace{\sc ii}] emission shouldn't be, at least for the bin at 11.5$'$. 

For the reasons mentioned above, we make the assumption that the extended diffuse gas we detected is truly associated with the galaxy, and that the [N\thinspace{\sc ii}] emission we see at 6566.4 \AA\ is the result of a phase jump, making its velocity ($\sim400$ km s$^{-1}$) consistent with the H$\alpha$ velocity field.

The study by \citet{Ken2008178} found a mean [N\thinspace{\sc ii}]/H$\alpha$ ratio of 0.48 for NGC 253, which includes the [N\thinspace{\sc ii}] lines at 6548 \AA\ and 6583 \AA . However, many studies have shown that this ratio can significantly change in areas located outside of dense H\thinspace{\sc ii} regions and where the DIG component is still present \citep{Rey1995448,Ran1997474}, i.e., at radii larger than the neutral component for NGC 253. [N\thinspace{\sc ii}] is a forbidden line, and can therefore be seen in areas of diffuse gas where spontaneous emission can become important. \citet{Ran1997474} and \citet{Ran1998501} showed for the galaxy NGC 891 that the [S\thinspace{\sc ii}]/H$\alpha$, [N\thinspace{\sc ii}]/H$\alpha$ and even [O\thinspace{\sc iii}]/H$\alpha$ ratio increase significantly with increasing distance to the midplane. For the DIG component of the Milky Way, the forbidden lines of [N\thinspace{\sc ii}] and [S\thinspace{\sc ii}] were also found to have intensities with respect to the H$\alpha$ line of a few tenths to unity, and even larger \citep{Haf200981}. This is much greater than that observed in classical H\thinspace{\sc ii} regions. Hence, forbidden lines could be better probes of the extended disc. This could explain the detection of [N\thinspace{\sc ii}] at radii larger than 11.5$'$ rather than H$\alpha$.  

The following paragraphs compare the results found in this paper with those of \citet{Bla1997490}. Table \ref{a1_t3} shows the EM values for the diffuse H$\alpha$ and [N\thinspace{\sc ii}] lines, both for this study and that of Bland-Hawthorn et al. (1997). The error of the flux values of our study are estimated to be 30 per cent. The flux values for the diffuse emission in \citet{Bla1997490}'s paper are given in milli-Rayleigh (mR). The SHASSA survey, used to calibrate our data, adopts the conversion that 1R is equal to an EM of $\sim2$ cm$^{-6}$ pc. We have used this conversion throughout this paper. From this conversion, the \citet{Bla1997490} fluxes give the EM values shown in Table \ref{a1_t3}. The EMs found in this study are therefore consistent with those found by \citet{Bla1997490}. 

Bland-Hawthorn et al. (1997) used the TAURUS-2 FP interferometer at the 3.9m Anglo-Australian Telescope for their observations on NGC 253. Four fields, each of 5$'$ each, were observed on the receding side of the galaxy. They used the FP Staring technique, which gave a single, very deep spectrum of a diffuse source for each field of view. The spectrum in a 40 \AA\ band is spread out radially from the optical axis across the field. This means, that for the spectrum obtained at each field, an emission line seen at a wavelength is an emission line coming from a precise radius in the field. They were able to see the H$\alpha$ and [N\thinspace{\sc ii}] (6548 \AA) line, with a total of 10 velocity measurements on their rotation curve. Two fields were placed at extended radii - the first was exposed for 9 hr and led to a [N\thinspace{\sc ii}] velocity measurement at 10.5$'$ and a H$\alpha$ velocity measurement at 12$'$; the second was exposed for 5 hr and led to a [N\thinspace{\sc ii}] velocity measurement at 13.5$'$ and a H$\alpha$ velocity measurement at 15$'$. The velocities found for the very extended and diffuse emission are shown in Table \ref{a1_t3}, corrected for the systemic velocity and inclination found in this study.

In \citet{Bla1994437}, they demonstrated that the TAURUS-2 FP interferometer could reach an H$\alpha$ EM of 0.02 cm$^{-6}$ pc at the $3\sigma$ level in 6 hr across the full FSR (57 \AA , in their study). Our calculations gave a $1\sigma$ upper limit for the EM of $\sim0.03$ cm$^{-6}$ pc (see Section 4), and we observed the receding side of the galaxy for a total of 6 hours. Our study therefore was not as deep as that of \citet{Bla1994437}, but was deep enough to detect extended EMs and went further out. 

According to Table \ref{a1_t3} and the bottom-right panel of Fig.\ref{a1f5}, our velocity measurements are consistent with those of \citet{Bla1994437}, within the allowed error bars, except for two discrepancies. First, no H$\alpha$ emission was seen in our study at 15$'$, whereas \citet{Bla1994437} found emission at that radius (EM of 0.08 cm$^{-6}$ pc). This EM is below our $3\sigma$ detection and it is possible that our detection limits were not sufficient to allow this detection. 

Second, our [N\thinspace{\sc ii}] velocity measurement at $12.8'\pm0.7'$ is lying below the rotation curve found by Bland-Hawthorn et al. (1997), and more precisely below the [N\thinspace{\sc ii}] velocity measurement of Bland-Hawthorn et al. (1997) at 13.5$'$. The disagreement between the velocities arises from different radial velocity measurements, with a $\sim20$ km s$^{-1}$ difference. Although the observations are made with a Fabry-Perot in both cases, the technique is not at all the same since there is no scanning in Bland-Hawthorn et al. (1997) observations. This enables to detect faint fluxes of emission lines with much more confidence but is less accurate for measuring the velocities since inhomogeneities in the galaxy (such as knots) can bias the result because they can show up even if their position in the field is not the right one for producing a constructive interference pattern. As a result it seems that the error bars on the [N\thinspace{\sc ii}] velocity points of Bland-Hawthorn et al. (1997) are optimistic and underestimated. It is probable that the right value of the rotation velocity of NGC253 at about 13$'$ radius is intermediate between the value found by Bland-Hawthorn et al. (1997) and our value, say around 210 km/s, hence in continuity with the H$\alpha$ values found by Bland-Hawthorn et al. (1997) at 12$'$ and 15$'$ radius.

\subsubsection{\normalsize Origin of the very extended DIG}

\citet{Hoy196316} mentioned for the first time the existence of a diffused matter of free electrons in the vicinity of the Galactic plane using a low frequency synchrotron survey. Since then, many new properties of the DIG have been discovered. Some of these properties include the high ratios of forbidden lines with respect to H$\alpha$ \citep{Ran1997474,Ran1998501}. Furthermore, it was found that several edge-on less active galaxies show a correlation between the DIG in the halo and the star formation rate (SFR) \citep{Ros2003406a,Ros2003406b,Mil2003148}. The study by \citet{Ros2003406a} also suggested that there was a minimum SFR per area required to mimic the disc-halo interaction. However, a more recent study from \citet{Oey2007661} concluded from their SINGS H$\alpha$ Survey (Survey for Ionization in Neutral Gas Galaxies) that the DIG does not seem to correlate neither with the Hubble type nor with the total SFR of the galaxy. 

Nevertheless, many theories were proposed to explain the existence of the DIG. \citet{Bla1997490} analysed the case of NGC 253. Several ionization and heating sources for the DIG were examined, including the following: metagalactic ionizing background, compact halo sources, compact disc sources (white dwarf population and hot horizontal-branch stars), ram pressure heating resulting from shock waves, exotic heat sources such as turbulent MHD waves, young stellar disc, dilute photoionization, gas-phase depletion from grains, Lyman continuum (Lyc) ionizing photons escaping from OB stars. However, the only possible source that could explain the observations was the one regarding hot young stars in the central regions of the galaxy looking out towards the outer part of the warped disc. Ionizing photons could escape from these stars through dust scattering, and would be more easily exposed to the outer regions because of the warped structure. This explanation is in agreement with the detection of both UV emission (probing hot stars) out to $r=13'$ (see Section 5) and diffuse ionized emission at $r=11.5'$, 12.8$'$, 19.0$'$.

The calculations by \citet{Bla1997490} also ruled out the source from the metagalactic ionizing background. Their study determined that the metagalactic ionizing background was around five times less than that required to explain the H$\alpha$ observations. Their calculations also showed that the photoionization model, accounting for the Lyc photons escaping from OB stars, could not account for the observations, since the observed ratio of [N\thinspace{\sc ii}] ($\lambda$ 6548 \AA )/H$\alpha$ was slightly higher than unity. This implies that the ratio of [N\thinspace{\sc ii}] ($\lambda$ 6583 \AA )/H$\alpha$ was close to three. The latter is twice as high as what was predicted by the photoionization model. The photoionization model also fails to explain the lack of He\thinspace{\sc i} ($\lambda$ 5876 \AA ) emission. Studies show that the ionizing flux necessary to explain the observed [N\thinspace{\sc ii}] and [S\thinspace{\sc ii}] of the DIG should allow the existence of the He\thinspace{\sc i} line \citep{Dom1994428}. 

Other studies have also shown a correlation between the DIG in the haloes of galaxies and their synchrotron radio-continuum. This might suggest that the DIG is correlated with cosmic rays and magnetic fields \citep*[see][]{Dah1995444}. However, \citet{Sci1995276} discussed that it was difficult for cosmic rays to penetrate clouds of gas, and that they should not play an important role in generating the DIG. Nevertheless, the Symposium paper by Heesen et al. (2009)\nocite{Hee2009259b} found a total equipartition magnetic field on the order of 7 $\mu$G to 18 $\mu$G for the disc of NGC 253. These values remain high. The magnetic field of NGC 253 could therefore play a key role in understanding the DIG component. The presence of a disc wind in NGC 253 was also discussed in the paper by Heesen et al. (2009)\nocite{Hee2009494a}, and could explain the presence of extra-planar H\thinspace{\sc i}, H$\alpha$, and soft X-ray emission in the halo. Finally, it is also important to note that the Magellanic streams can not explain the [N\thinspace{\sc ii}] emission found in this study, since the radial velocities of the two do not correlate spatially \citep{Mat1984108}.

\subsection{\normalsize Consequences of a declining rotation curve}\label{a1c63}

Declining rotation curves in spiral galaxies are rare. However, some have been seen in earlier-type systems \citep{Cas1991101}, as well as in the edge-on Sc galaxy NGC 891 \citep{San197974}. The Gassendi H$\alpha$ survey of SPirals (GHASP) also find a sub-sample of galaxies showing rotation curves slightly declining \citep[see][]{Gar2002387,Gar2004349,Gar2005362,Epi2008388}. Moreover, \citet{Dic2008135} find, for the late-type system NGC 7793, a decline of $\sim25$ per cent $v_\mathrm{max}$ over half the radius range. For NGC 253, \citet{Bla1997490} find a decline at the last observed point of the rotation curve of $\sim10$ per cent $v$$_\mathrm{max}$ over the last $\sim10$ per cent of $r_\mathrm{max}$. This study finds a decline of $\sim30$ per cent $v_\mathrm{max}$ over slightly more than half the radius range. The only way not to have a declining rotation curve would be if the inclination was smaller by 30$^o$ for the last observed velocity point, which is quite unlikely (see Fig. \ref{a1f5}).

The top panels of Fig. \ref{a1f6a}, Fig. \ref{a1f6b} and Fig. \ref{a1f6c} show the results for the maximum stellar disc model, respectively for the H\thinspace{\sc i} data of \citet{Puc1991101}, the H$\alpha$ data without the DIG points, and the H$\alpha$ data with the DIG points. In all cases, the observed rotation curve can be reproduced by a stellar disc and a gaseous disc. Hence, no dark halo seems to be needed. However, in the central regions of the H$\alpha$ rotation curve, the presence of a bulge can be observed as seen by the \textit{bump} in the curve. To reproduce well this region, a bulge component could be added to the maximum stellar disc model. 

An advantage of maximum disc models is that a unique solution can be determined, whereas many degeneracies exist in multicomponent fits including dark haloes \citep{Car1985294,vanA1986320}. Maximum stellar discs also seem to reproduce well the small scale irregularities in the rotation curves, since the stellar disc also traces these perturbations \citep{Fre1992}. However, as can be seen in Table \ref{a1_t4}, the (\textit{M}/\textit{L})$_\star$ that best fits the maximum-disc model is of $\sim12$ M$_\odot$/L$_\odot$ for all three cases. This value is much too high, especially in the case of late-type galaxies such as NGC 253, which causes great concern for the validity of the model. 

The study by \citet{Cou1999513} showed that the Tully-Fisher relation (TFR) seemed to be independent of surface brightness for high surface brightness (HSB) galaxies when plotted as$M_\mathrm{r}$ versus $V_{2.2}$ ($V_{2.2}$ is the velocity for an exponential disc at 2.2 scale lenghts). In order to reproduce this feature, their models suggested that discs were sub-maximum with $v_\mathrm{disc}/v_\mathrm{total}\sim0.6$. Additionally, their study suggested that these sub-maximum discs allowed a smooth transition between HSB and low surface brightness galaxies, and that maximum discs for the majority of the HSB late-type galaxies are not valid. 

The bottom of Fig. \ref{a1f6a}, Fig. \ref{a1f6b} and Fig. \ref{a1f6c} show the results for sub-maximum discs, respectively for the H\thinspace{\sc i} data of \citet{Puc1991101}, the H$\alpha$ data without the DIG points, and the H$\alpha$ data with the DIG points. For NGC 253, the (\textit{M}/\textit{L})$_\star$ was fixed to 5.0 M$_\odot$/L$_\odot$ to allow $v_\mathrm{disc}/v_\mathrm{total}\sim0.6$. For all three cases, the sub-maximum disc seems to agree well with the derived rotation curves. The decline observed with the DIG points is also well reproduced. The model also yields very similar dark halo parameters for the rotation curve with and without the DIG component. Hence, the derived rotation curve for NGC 253 using the DIG points truly seems to illustrate a declining rotation curve, and hints that the end of the mass distribution has started to be probed. 

Moreover, the sub-maximum disc gives a steeper slope in the central areas which seems to reproduce better the H$\alpha$ data at these radii. Table \ref{a1_t4} shows that the H$\alpha$ data have a higher central dark halo density than the H\thinspace{\sc i} data, which can be explained by the steeper slope of the H$\alpha$ data. However, the dark halo velocity dispersion $\sigma$ seems to be fairly the same in all three cases. This parameter is a measure of the maximum amplitude of the rotation curve \citep{Bla1999118}, and since the maximum velocity is essentially the same for the H\thinspace{\sc i} and H$\alpha$ data, similar values are expected. Finally, the (\textit{M}/\textit{L})$_\star$ in the \textit{B} band expected from stellar population models, using the (\textit{B}-\textit{V}) given by the RC3 catalogue (0.85 mag), is $\sim3.46$ \textit{M}$_\odot$/\textit{L}$_\odot$ \citep{Bel2001550}. This is much more consistent with the ratio used for the sub-maximum disc than the one determined from the maximum disc. 
\begin{figure}
\centering
\begin{minipage}[c]{0.99\linewidth}
 \centering \includegraphics[width=\linewidth]{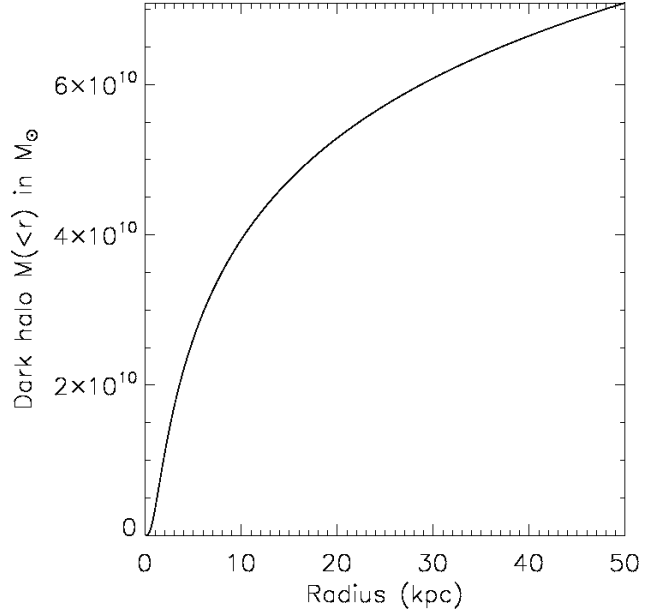}
\end{minipage}
\caption{Cumulative mass profile ($M(<r)$) of the dark halo. An isothermal sphere was assumed for the dark halo, with $\rho$$_\mathrm{0}=1.564$ M$_\odot$ pc$^{-3}$ and $r_\mathrm{c}=1.0$ kpc. These are the best fit parameters of the dark halo for the sub-max disc fit to the data that include the diffuse emission. The total mass is in units of M$_\odot$. }
\label{a1f8}
\end{figure}

Our results supports Bland-Hawthorn et al. (1997)'s suggestion that the rotation curve is declining in the outer parts.  While their data points were showing a decline between 8 and 9.5 kpc, our outermost velocity point shows that the decline seems to continue out to $\sim14$ kpc. As seen in the mass model analysis, there are two possible interpretations: 1- either the halo is truncated or 2- the halo is highly concentrated (see fig. 7, with $r_\mathrm{c}$ $\sim$ 1 kpc). Although the second interpretation is not favoured in view the mass distribution usually seen in late-type spirals were $r_\mathrm{c}$ is more of the order of 5-10 kpc (see e. g. Spano et al. 2008)\nocite{Spa2008383}, it seems to be the right one 
since Fig. \ref{a1f8} (cumulative mass curve of the dark halo as a function of
radius) shows that the halo is not truncated since it is continuously rising. This profile was derived using the parameters found with the extended H$\alpha$ rotation curve.

NGC 253 is the second late-type spiral in the Sculptor group with NGC 7793 (Dicaire et al. 2008) that shows a declining rotation curve in the outer parts. For NGC 253 and NGC 7793, the halo maximum radius seem to be $\sim10$ kpc. This is smaller than the maximum radius of $\sim40$ kpc suggested by a dynamical analysis of the Sculptor Group (Puche \& Carignan 1991). However, this study was assuming that the Sculptor group was virialized which is not necessarily the case since the crossing time is of the same order as a Hubble time. If the group is not virialized, their estimate would then be an upper limit.

\section{\large{Conclusion}}

Deep FP H$\alpha$ observations were obtained for one of the Sculptor Group galaxies, NGC 253. The main conclusions of this study are as follows: 

1. The velocity field was obtained using three methods. The first lead to an initial velocity field reaching a distance of 11.5$'$. The second, in order to extract the global kinematical parameters, used a gaussian smoothing function ($w_\mathrm{\lambda}=55''$) and lead to a velocity field reaching 10$'$. The third used a pie--shape smoothing, in order to extract the profiles of the DIG component. Typical flux values of $\sim0.1$ cm$^{-6}$ pc were reached. 

2. The nuclear region showed the presence of a bar-like structure lagging in velocity and showing an increase in dispersion. 

3. The observations allowed the detection of the DIG component through H$\alpha$ emission at a radius of 11.5$'$ and [N\thinspace{\sc ii}] emission at radii of 11.5$'$, 12.8$'$ and 19.0$'$, on the receding side of the galaxy. No H$\alpha$ emission was observed at radii larger than the neutral component (11.5$'$).

4. From the gaussian smoothed velocity field, the global kinematical parameters were derived and yielded a mean systemic velocity of $241.2\pm5.4$ km s$^{-1}$, an inclination \textit{i} of $76.4\pm3.9^o$ and a position angle PA of $230.1\pm2.1^o$. A warp is also seen through the variation of the kinematical parameters with radius. 

5. A rotation curve was extracted using those global kinematical parameters and the initial velocity field. This rotation curve is in good agreement with the H\thinspace{\sc i} curve derived by \citet{Puc1991101}, although the slope of the inner region is steeper (as expected from the beam smearing of the H\thinspace{\sc i} data). The data points resulting from the extended diffuse gas (H$\alpha$; [N\thinspace{\sc ii}] ($\lambda$ 6548 \AA ) were added and showed a significant decline although slightly different from the results of \citet{Bla1997490}. 

6. A mass model analysis was obtained for the H$\alpha$ rotation curve with and without the velocities derived for the extended DIG emission. The two yielded very similar results, both for the maximum stellar disc and the sub-maximum stellar model. The declining part of the rotation curve was also very well modeled, and seems to truly be declining. 

Deep observations of galaxies can yield very interesting results, as was the case for the study by \citet{Dic2008135} which allowed confirmation of a truly declining rotation curve of NGC 7793. In this study, the emission detected in NGC 253 of the DIG confirmed the declining nature of the rotation curve. This has profound consequences on the study of dark matter properties, and could allow very good constraints of the parameters of the dark halo. 

\section*{Acknowledgments}

We would like to thank the staff from the ESO La Silla site, in Chile, for their support, as well as the referee for all of his/her useful comments. Moreover, we would like to thank Monica Relano-Pastor for taking time from her busy schedule to pre-read this article, as well as thank the students from the 1.2-m Euler Telescope for lending us their computer during the observation mission. We recognize all the support given by the Natural Sciences and Engineering Research Council of Canada, as well as the Fonds Québécois de la Recherche sur la Nature et les Technologies. The Southern H-Alpha Sky Survey Atlas (SHASSA), which is supported by the National Science Foundation, was used for flux calibration of the data in this study. The images were taken with a robotic camera operating at Cerro Tololo Inter-American Observatory (CTIO), in Chile. 

\label{lastpage}

\bibliographystyle{mn2e}
\bibliography{253bib}

\end{document}